\documentclass[sn-mathphys]{sn-jnl}% Math and Physical Sciences Reference Style
\jyear{2023}%

\usepackage{graphicx}
\usepackage{lscape}
\usepackage{adjustbox}
\usepackage{bm}
\usepackage{cleveref}
\usepackage{amssymb}
\usepackage{psfrag}
\usepackage{color}
\usepackage{amsbsy}      
\usepackage{amsmath}
\usepackage{mathtools}
\usepackage{dsfont}
\usepackage{lineno}
\usepackage{comment}
\usepackage{hyperref}
\usepackage{hhline}
\usepackage{float,lscape}
\usepackage{url}
\usepackage{soul,color}

\crefname{equation}{}{}
\def\ie{\textit{i.e.}}
\def\eg{\textit{e.g.}}

\def\dd{\textrm{d}}

\def\p{{\bf p}}

\def\n{{\bf n}}
\def\s{{\bf s}}

\begin{document}

\title[Private and Public TCR statistics]{Mathematical
  Characterization of Private and Public Immune Receptor Sequences}


\author[1,2]{\fnm{Lucas} \sur{B\"ottcher}}\email{l.boettcher@fs.de}

\author[3]{\fnm{Sascha} \sur{Wald}}\email{sascha.wald@coventry.ac.uk}

\author*[2,4]{\fnm{Tom} \sur{Chou}}\email{tomchou@ucla.edu}

\affil[1]{\orgdiv{Dept. of Computational Science and Philosophy},
  \orgname{Frankfurt School of Finance and Management},
  \orgaddress{\street{Frankfurt am Main}, \city{Frankfurt},
    \postcode{60322}, \country{Germany}}}

\affil*[2]{\orgdiv{Dept.~of Computational Medicine}, \orgname{University of California, Los Angeles},
  \orgaddress{\street{621 Charles E. Young Dr. S.}, \city{Los Angeles}, \postcode{90095-1766}, \state{California},
    \country{United States}}}

\affil[3]{\orgdiv{Statistical Physics Group, Centre for Fluid and Complex Systems}, \orgname{Coventry University},
  \orgaddress{\street{Priory Street}, \city{Coventry}, \postcode{CV1 5FB}, \country{United Kingdom}}}

\affil[4]{\orgdiv{Dept.~of Mathematics}, \orgname{University of California, Los Angeles},
  \orgaddress{\street{520 Portola Plaza}, \city{Los Angeles}, \postcode{90095-1555}, \state{California},
    \country{United States}}}

% REQUIRED
\abstract{Diverse T and B cell repertoires play an important role in
  mounting effective immune responses against a wide range of
  pathogens and malignant cells. The number of unique T and B cell
  clones is characterized by T and B cell receptors (TCRs and BCRs),
  respectively. Although receptor sequences are generated
  probabilistically by recombination processes, clinical studies found
  a high degree of sharing of TCRs and BCRs among different
  individuals. In this work, we use a general probabilistic model for
  T/B cell receptor clone abundances to define ``publicness'' or
  ``privateness'' and information-theoretic measures for comparing the
  frequency of sampled sequences observed across different
  individuals. We derive mathematical formulae to quantify the mean
  \textit{and the variances} of clone richness and overlap.  Our
  results can be used to evaluate the effect of different sampling
  protocols on abundances of clones within an individual as well as
  the commonality of clones across individuals.  Using synthetic and
  empirical TCR amino acid sequence data, we perform simulations to
  study expected clonal commonalities across multiple
  individuals. Based on our formulae, we compare these simulated
  results with the analytically predicted mean and variances of the
  repertoire overlap. Complementing the results on simulated
    repertoires, we derive explicit expressions for the richness and
    its uncertainty for specific, single-parameter truncated power-law
    probability distributions. Finally, the information loss
  associated with grouping together certain receptor sequences, as is
  done in spectratyping, is also evaluated. Our approach can be, in
  principle, applied under more general and mechanistically realistic
  clone generation models.}




% REQUIRED
%\begin{AMS}
%62P10, 92C37, 94A16
%\end{AMS}
%
\date{\today}

\maketitle

\section{Introduction}

A major component of the adaptive immune system in most jawed
vertebrates is the repertoire of B and T lymphocytes.  A diverse
immune repertoire allows the adaptive immune system to recognize a
wide range of pathogens~\cite{xu2020diversity}. B and T cells develop
from common lymphoid progenitors (CLPs) that originate from
hematopoietic stem cells (HSCs) in the bone marrow. B cells mature in
the bone marrow and spleen while developing T cells migrate to the
thymus where they undergo their maturation process. After encountering
an antigen, naive B cells may get activated and differentiate into
antibody-producing plasma cells, which are essential for humoral (or
antibody-mediated) immunity. In recognizing and eliminating infected
and malignant cells, T cells contribute to cell-mediated immunity of
adaptive immune response.

T-cell receptors bind to antigenic peptides (or epitopes) that are
presented by major histocompatibility complex (MHC) molecules on the
surface of antigen-presenting cells (APCs). T cells that each carry a
type of TCR mature in the thymus and undergo V(D)J recombination,
where variable (V), diversity (D), and joining (J) gene segments are
randomly recombined~\cite{alt1992vdj,travers1997immunobiology}. The
receptors are heterodimeric molecules and mainly consist of an
$\alpha$ and a $\beta$ chain while only a minority, about
1--10\%~\cite{girardi2006immunosurveillance}, of TCRs consists of a
$\delta$ and a $\gamma$ chain. The TCR $\alpha$ and $\gamma$ chains
are made up of VJ and constant (C) regions. Additional D regions are
present in $\beta$ and $\gamma$ chains. During the recombination
process, V(D)J segments of each chain are randomly recombined with
additional insertions and deletions.  After recombination, only about
5\% or even less~\cite{yates2014theories} of all generated TCR
sequences are selected based on their ability to bind to certain MHC
molecules (``positive selection'') and to not trigger autoimmune
responses (``negative selection''). These naive T cells are then
exported from the thymus into peripheral tissue where they may
interact with foreign peptides that are presented by APCs. The selection
process as well as subsequent interactions are specific to an
individual.

The most variable parts of TCR sequences are the complementary
determining regions (CDRs) 1, 2, and 3, located within the V region,
among which the CDR3$\beta$ is the most
diverse~\cite{abbas2021cellular}. Therefore, the number of distinct
receptor sequences, the richness $R$, of TCR repertoires is typically
characterized in terms of the richness of CDR3$\beta$ sequences. Only
about 1\% of T cells express two different TCR$\beta$
chains~\cite{davodeau1995dual,padovan1995normal,schuldt2019dual},
whereas the proportion of T cells that express two different
TCR$\alpha$ chains may be as high as
30\%~\cite{rybakin2014allelic,schuldt2019dual}.

B cells can also respond to different antigens via different B cell
receptors (BCRs) that are comprised of heavy and light chains.  As
with TCRs, the mechanism underlying the generation of a diverse pool
of BCRs is VDJ recombination in heavy chains and VJ recombination in
light chains. Positive and negative selection processes sort out about
90\% of all BCRs that react too weakly or strongly with certain
molecules~\cite{tussiwand2009tolerance}. As a result of the various
recombination and joining processes and gene insertions and deletions,
the \textbf{practical theoretical maximum repertoire size $\Omega_{0}$} of the
variable region of BCR and TCR receptors \textbf{can be} $\sim
10^{14}-10^{20}$~\cite{davis1988t,venturi2008molecular,zarnitsyna2013estimating,lythe2016many}. This
value is comparable to the possible number of amino acid sequences of
typical length $\sim 11-12$. However, many of these sequences are not
viable, \textbf{are removed through thymic selection, or are have such
  low probability occuring that they are never expected to be  produced in an
  organism's lifetime.}  Thus, the effective number
of TCR variable regions that are produced and that can
\textbf{contribute to the organism's repertoire size, $\Omega$, should} be much
less than $\Omega_{0}$.  Estimating the true \textbf{size} of BCR and TCR
repertoires realized in an organism is challenging since the majority
of such analyses are based on small blood samples, leading to problems
similar to the ``unseen species'' problem in
ecology~\cite{laydon2015estimating}.  Nonetheless, the number of
unique TCRs realized in organisms has been estimated to be about
$10^6$ for mice~\cite{casrouge2000size} and about $10^8$ for
humans~\cite{soto2020high}.  B-cell \textbf{repertoire size} for humans is estimated
to be $10^8-10^9$~\cite{dewitt2016public}. These values are
significantly smaller than \textbf{$\Omega_{0}$ and might be used
as an effective $\Omega$.}

Each pool of BCR and TCR sequences realized in one organism $i$ can be
seen as a subset $\mathcal{U}_i$ of the set of all possible
species-specific sequences $\mathcal{S}$. Sequences that occur in at
least two different organisms $i$ and $j$ (\ie, sequences that are
elements of $\mathcal{U}_i\cap \mathcal{U}_j$) are commonly referred
to as ``public'' sequences~\cite{laydon2015estimating} while
``private'' sequences occur only in one of the individuals tested. The
existence of public TCR$\beta$ sequences has been established in
several previous
works~\cite{putintseva2013mother,robins2010overlap,shugay2013huge,soto2020high}. More
recently, a high degree of shared sequences has been also observed in
human BCR repertoires~\cite{briney2019commonality,soto2019high}.

The notions of public and private clonotypes have been loosely
defined. Some references use the term ``public sequence'' to refer to
those sequences that ``are \emph{often} shared between
individuals''~\cite{shugay2013huge} or ``shared across
individuals''~\cite{greiff2017learning}. \textbf{Recently, Elhanati et
  al.~\cite{elhanati2018predicting} and \cite{ruiz2023modeling} have
  formulated mathematical and statistical framework to quantify
  ``publicness'' and ``privateness.'' Building on these works, we
  derive a set of measures that enable us to quantify immune
  repertoire properties, including the expected total richness, the
  expected numbers of public and private clones, \emph{and their
    variances} (confidence levels), all expressed in terms of the
  general set of clone generation probabilities or clone populations.
  One of our metrics is the expected ``$M$-overlap'' or
  ``$M$-publicness,'' defined as the expected number of clones that appear in
  samples drawn from $M$ separate individuals. This quantity is a clinically
  interpretable limit of the expected ``sharing number'' defined in
  Elhanati et al.~\cite{elhanati2018predicting}. Similarly, we define
  $M$-private clones as clones that are not shared by all $M$
  individuals, \ie, occurring in at most $M-1$ individuals.}

In the next section, we first give an overview of the mathematical
concepts that are relevant to characterize TCR and BCR
distributions. We then formulate a statistical model of receptor
distributions in Sec.~\ref{sec:stat_model}. In
Secs.~\ref{sec:single_organism} and \ref{sec:collective_quantities},
we derive quantities associated with receptor distributions in single
organisms and across individuals, respectively. We will primarily
focus on the overlap of repertoires across individuals and on the
corresponding confidence intervals that can be used to characterize
``public'' and ``private'' sequences of immune repertoires.  Formulae
we derived are listed in Table \ref{tab:results}. In
Sec.~\ref{sec:simus}, we use synthetic and empirical TCR amino acid
sequence data and perform simulations to compare theoretical
predictions of repertoire overlaps between different individuals with
corresponding observations. Finally, when analyzing empirical
  sequence data, one may use continuous
  approximations~\cite{elhanati2018predicting,ruiz2023modeling} and
  averaging (\ie, coarse-graining) methods that change the information
  content in the underlying dataset. Coarse-graining of TCR and BCR
  data may also be a result of the employed sequencing techniques
  \cite{gorski1994circulating,fozza2017study}. In
  Sec.~\ref{sec:information_loss}, we therefore briefly discuss the
  information loss associated with analyzing processed cell data. We
  discuss our results and conclude our paper in
  Sec.~\ref{sec:conclusion}. Our source codes are publicly available
at \cite{GitLab}.

\section{Mathematical concepts}
\label{sec:math_concepts}
\begin{figure}
\centering
\includegraphics[width=0.94\textwidth]{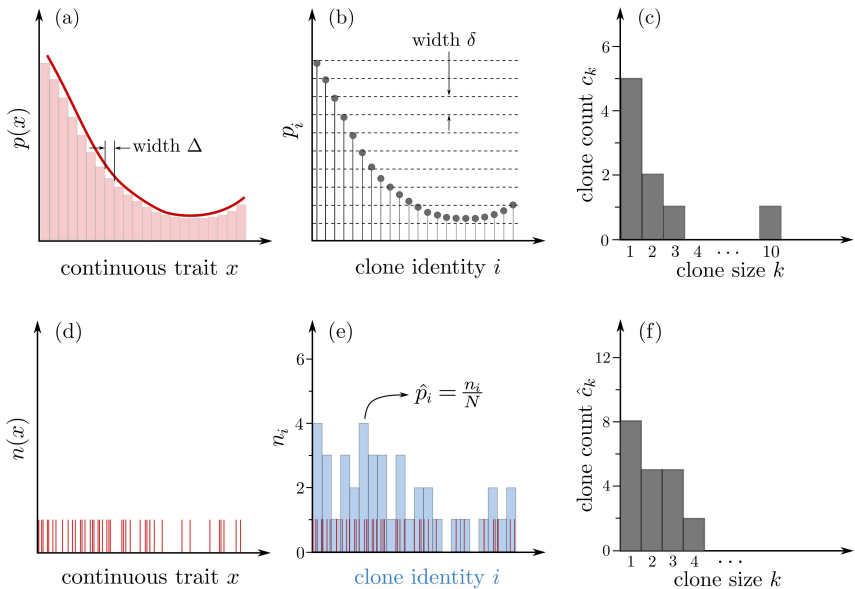}
\caption{Sampling from a continuous distribution, described in terms
  of an underlying probability density $p(x)$ and number density
  $n(x)$.  The probability density $p(x)$ (solid red line) and the
  Riemann sum approximation to the probability (red bars of width
  $\Delta$) are shown in panel (a).  The probability that a trait in
  the interval $[x,x+\mathrm{d}x)$ arises is $p(x)\mathrm{d}x$. As
    shown in (b), this distribution can be discretized directly by the
    intervals $[i\Delta,(i+1)\Delta)$ (red bars) defining discrete
      traits and their associated probabilities $p_i$ (see
      Eq.~\eqref{eq:f_i}). The probabilities $p_i$ can be transformed
      into clone counts $c_k$ (the number of identities $i$ that are
      represented by $k$ individuals) using
      Eq.~\eqref{eq:clone_count_cont}, which are shown in (c). A finite
      sample of a population described by $p(x)$ yields the binary
      outcome shown in (d).  In this example, the total number of
      samples is $N=41$ and and since the trait space $x$ is
      continuous, the probability that the exact same trait arises in
      more than one sample is almost surely zero. Light blue bars in
      panel (e) represent number counts $n_i$ binned according to
      $\Delta$.  The probabilities $\hat{p}_i=n_i/N$ provide an
      approximation of $p_i$. Clone counts for the empirical
      $\hat{p}_i$ are calculated according to
      Eq.~\eqref{eq:clone_count} and shown in (f).}
\label{CONFIGS}
\end{figure}
Although receptor sequences and cell counts are discrete quantities,
using continuous functions to describe their distribution may
facilitate the mathematical analysis of the quantities that we derive
in the subsequent sections. For example, a continuous approximation
(\ie, a ``density-of-states approximation'') of receptor sequence
distributions has been used in
\cite{elhanati2018predicting,ruiz2023modeling} to analyze properties
of immune repertoires. We therefore briefly review some elementary
concepts associated with continuous distributions and their
discretization.

Let $p(x)$ be the {\it probability density} associated with the
distribution of traits, as depicted in Fig.~\ref{CONFIGS}(a). The
probability that a certain trait occurs in $[x,x+\mathrm{d}x)$ is
  $p(x)\,\mathrm{d}x$. The corresponding discretized distribution
  elements are
\begin{equation}
p_i\coloneqq\int_{i \Delta}^{(i+1)\Delta}\!\!\! p(x)\,\mathrm{d}x,
\label{eq:f_i}
\end{equation}
where $\Delta$ is the discretization step size of the support of
$p(x)$. If we discretize the values of probabilities, the number of
clones with a certain relative frequency $p_i$ is given by the {\it
  clone count}
\begin{equation}
c_k\coloneqq \sum_i \mathds{1}\big(k\delta \leq p_{i}<(k+1)\delta\big),
  %c_k\coloneqq \sum_i \Theta[f_i-k\delta]\Theta[(k+1)\delta-p_i],
\label{eq:clone_count_cont}
\end{equation}
where the indicator function $\mathds{1}=1$ if its argument is
satisfied and 0 otherwise. As shown in Fig.~\ref{CONFIGS}(b), the
parameter $\delta$ defines an interval of frequency values and
modulates the clone-count binning.  Figures~\ref{CONFIGS}(b,c) show
how $p_i$ and $c_k$ are constructed from a continuous distribution
$p(x)$.

If $p(x)$ is not available from data or a model, an alternative
representative starts with the number density $n(x)$, which can be
estimated by sampling a process which follows $p(x)$.  The probability
that a continuous trait $x$ is drawn twice from a continuous
distribution $p(x)$ is almost surely zero. Hence, the corresponding
number counts $n(x)$ are either 1 if $X\in[x,x+\mathrm{d}x)$ (\ie, if
  trait $X$ is sampled) or 0 otherwise, as shown in
  Figs.~\ref{CONFIGS}(d,e). We say that $X$ is of \textit{clonotype}
  $i$ if $X\in [i \Delta,(i+1)\Delta)$ ($1\leq i \leq \Omega$) and we
    use $n_i$ to denote the number of cells of clonotype $i$. Then, if
    $\Omega$ denotes the effective number of different clonotypes, the
    total T-cell (or B-cell) population is $N\equiv \sum_{i=1}^\Omega
    n_i$.  The relative empirical abundance of clonotype $i$ is thus
    $\hat{p}_i=n_i/N$ (see Fig.~\ref{CONFIGS}e), satisfying the
    normalization condition $\sum_i \hat{p}_i=1$. The corresponding
    empirical clone count derived from the number representation
    $n_{i}$ is defined as
\begin{equation}
\hat{c}_{k} \coloneqq \sum_{i=1}^{\Omega}\mathds{1}(n_{i},k)
\label{eq:clone_count}
\end{equation}
and shown in Fig.~\ref{CONFIGS}(f). \textbf{The indicator function
  $\mathds{1}(a,b)$ with arguments $a,b\in \mathds{Z}_{\geq 0}$ is
  equal to 1 if $a=b$ and 0 otherwise.} Clone counts can also be used
to describe T cell repertoires, especially if clone identities are not
important. Simple birth-death-immigration models can also be cast in
terms of, \textit{e.g.}, expected clone counts
$\mathbb{E}[\hat{c}_{k}(t)]$ \cite{Goyal2015,Lewkiewicz2019}. Besides
the simple discrete estimate $\hat{p}_i=n_i/N$, one can also
reconstruct $p(x)$ from $\n=\{n_{i}\}$ using methods such as kernel
density estimation.

%The sum in Eq.~\eqref{eq:clone_count} counts the
%number of clonotypes with the same clonotype number. For example, if
%$c_5=2$ there are two clones with abundance $n = 5$.

\section{Whole organism statistical model}
\label{sec:stat_model}
Using the mathematical quantities defined above, we develop a simple
statistical model for BCR and TCR sequences distributed among
individuals. Although our model is applicable to both BCR
and TCR sequences, we will primarily focus on the characterization of
TCRs for simplicity. B cells undergo an additional
process of somatic hypermutation and class switching leading to a more
dynamic evolution of the more diverse B cell repertoire
\cite{elhanati2015inferring}. By focusing on naive T cells, we can
assume their populations are generated by the thymus via a single,
simple effective process.

\begin{figure}[htb]
    \centering
    \includegraphics[width=0.92\textwidth]{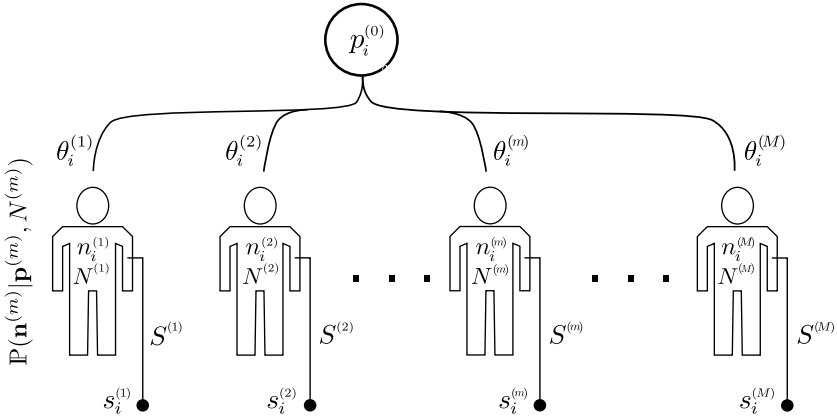}
    \caption{Schematic of sampling of multiple species from multiple
      individuals.  A central process produces (through V(D)J
      recombination) TCRs. Individuals select for certain TCRs
      resulting in population $n_{i}^{(m)}$ of T cells with receptor
      $i$ in individual $m$, for a total T-cell count $N^{(m)} =
      \sum_{i}^{\Omega}n_{i}^{(m)}$. The selection of TCR $i$ by
      individual $m$ (in their individual thymuses) is defined by the
      parameter $\theta_i^{(m)}$ which gives an effective probability
      $p_{i}^{(m)}\equiv \theta_{i}^{(m)} p_{i}^{(0)}$. A sample with
      cell numbers $S^{(m)} \ll N^{(m)}$ is drawn from individual $m$
      and sequenced to determine $s_{i}^{(m)}$, the number of cells of
      type $i$ in the subsample drawn from individual $m$.}
    \label{FIG2}
\end{figure}

Assume a common universal recombination process (see Fig.~\ref{FIG2})
in T-cell development that generates a cell carrying TCR of type
$1\leq i\leq \Omega_{0}$ with probability $p_{i}^{(0)}$. Here,
$\Omega_{0}\gg \Omega$ is the the theoretical number of ways the full
TCR sequence can be constructed which is itself much larger than the
\textbf{effective} number $\Omega$ that appears in an individual after
thymic selection.
%
% The theoretical number of possible sequences before thymic selection
% is very large, $\Omega \sim 10^{16}$.
%
Although each new T cell produced carries TCR $i$ with probability
$p_{i}^{(0)}$, many sequences $i$ are not realized given thymic
selection (that eliminates $\sim 98\%$ of them), the finite number of
T cells produced over a lifetime
\cite{travers1997immunobiology,yates2014theories,lythe2016many}, or
the extremely low generation probability of some clones. These effects
are invoked to truncate $\Omega_{0}$ to $\Omega\ll \Omega_{0}$.
However, we will see in Section \ref{sec:forms}, explicit scaling
relastionships for the limit $\Omega \gg 1$ can be found for general
power-law ordered probabilities $p_{i}$.

\textbf{Besides VDJ recombination~\cite{slabodkin2021individualized},
  thymic selection and subsequent death, activation, and proliferation
  occur differently across individuals $1\leq m \leq M$ and may be described by
  model parameters $\theta_{i}^{(m)}$.} Such a model translates the
fundamental underlying recombination process into a population of
$n_{i}^{(m)}$ T cells with TCR $i$ and total population $N^{(m)}=
\sum_{i=1}^{\Omega}n_{i}^{(m)}$ in individual $m$.  The connection
between $p_{i}^{(0)}, \theta_{i}^{(m)}$ and $n_{i}^{(m)}, N^{(m)}$
might be described by dynamical models, deterministic or stochastic,
such as those presented in \cite{dessalles2021naive}.

At any specific time, individual $m$ will have a cell population
configuration $\n^{(m)} \equiv (n_{1}^{(m)}, n_{2}^{(m)}, \ldots,
n_{\Omega}^{(m)})$ with probability $\mathds{P}(\n^{(m)}\vert
\boldsymbol{\theta}^{(m)},N^{(m)})$. Each individual can be thought of
as a biased sample from all cells produced via the universal
probabilities $p_{i}^{(0)}$.  We can approximate individual
probabilities $p_{i}^{(m)} \equiv \theta_{i}^{(m)}p_{i}^{(0)},\, 1\leq
i \leq \Omega$, where the number of effective TCRs
  $\Omega$ for individual $m$ might have as upper bound
  $\Omega \sim 10^{14}$, if, for example, we are considering just
  the CDR3 region of the $\beta$ chain.  Assuming a time-independent
model (\textit{e.g}, a model in steady-state), we can describe the
probability of a T-cell population $\n^{(m)}$ in individual $m$ by a
multinomial distribution over individual probabilities $\p^{(m)}
\equiv \{p_{i}^{(m)}\}$:

\begin{equation}
\mathds{P}(\n^{(m)}\vert \p^{(m)},N^{(m)}) = 
N^{(m)}! \prod_{i=1}^{\Omega}{[p_{i}^{(m)}]^{n_{i}^{(m)}}\over
n_{i}^{(m)}!},
\label{MULTINOMIAL_n}
\end{equation}
where $\sum_{i=1}^{\Omega}n_{i}^{(m)} \equiv N^{(m)}$ and
$\sum_{i=1}^{\Omega}p_{i}^{(m)} \equiv 1$. Thus, each
  individual can be thought of as a ``sample'' of the ``universal''
  thymus.  Neglecting genetic relationships amongst individuals, we
  can assume them to be independent with individual probabilities
  $p_{i}^{(m)}$. These probabilities describe the likelihood that a
  randomly drawn cell from individual $m$ is a cell of clone $i$.
  Repeated draws (with replacement) would provide the samples for the
  estimator $\hat{p}_{i}^{(m)}= n_{i}^{(m)}/N^{(m)}$, assuming
  $n_{i}^{(m)}$ are counted and $N^{(m)}$ is also known or estimated.
This representation allows us to easily express the probabilities of
any configuration $\n^{(m)}$ analytically.  A dynamical model for
$n_{i}^{(m)}$ cannot be directly described by our simple probabilities
$p_{i}^{(m)}$. A mechanistically more direct model could incorporate
the production rate of clone $i$ T cells from the thymus, the
proliferation and apoptosis rates of clone $i$ cells, and interactions
manifested as, \textit{e.g.}, carrying capacity as model
parameters. Probability distributions for $\n^{(m)}$, as a function of
birth, death, and immigration rates, have been found in \cite{BDI} and
can also be used, instead of Eq.~\eqref{MULTINOMIAL_n}, to construct
probabilities.

%Key quantities to infer are the distribution of TCR $i$. For example,
%we may ask, in how many of the $M$ total individuals does TCR $i$
%appear?  

\subsection{Single individual quantities}
\label{sec:single_organism}
First, we focus on quantities intrinsic to a single individual
organism; thus, we can suppress the ``$m=1$'' label.  Within an
individual, we can use clone counts to define measures such as the
richness

\begin{equation}
R(\n) \coloneqq \sum_{i=1}^\Omega \mathds{1}(n_{i}\geq 1)  =\sum_{k \geq 1}
\sum_{i=1}^{\Omega} \mathds{1}(n_{i}, k) 
\equiv \sum_{k\geq 1}\hat{c}_{k},
\label{eq:richness_one}
\end{equation}
where $\hat{c}_{k}\equiv \sum_{i=1}^{\Omega} \mathds{1}(n_{i}, k)$ is
defined in Eq.~\eqref{eq:clone_count} (the number of clones that are
of size $k$).  Other diversity/entropy measures such as Simpson's
indices, Gini indices, etc.~\cite{rempala2013methods,xu2020diversity} can also be
straightforwardly defined. Given the clone populations $\n$, the
individual richness can be found by direct enumeration of
Eq.~\eqref{eq:richness_one}.

We can also express the richness in terms of the underlying
probabilities $\p$ associated with the individual by first finding the
probability $\rho_{i}$ that a type-$i$ cell appears at all
among the $N$ cells within said individual. This probability is

\begin{equation}
\rho_{i} \equiv \mathds{P}(n_{i}\geq 1\vert \p, N) = 1-(1-p_{i})^{N}
\label{rhoN}
\end{equation}
and corresponds to that of a binary outcome, either appearing or not
appearing.  Higher order probabilities like $\rho_{ij}$ (both $i$- and
$j$-type cells appearing in a specific individual) can be computed
using the marginalized probability

\begin{equation}
\mathds{P}(n_{i}, n_{j} \vert \p, N) 
= {N!\, p_{i}^{n_{i}}p_{j}^{n_{j}}
(1-p_{i}-p_{j})^{N-n_{i}-n_{j}}
\over n_{i}!\, n_{j}!\,  (N-n_{i}-n_{j})!}
\label{ninj}
\end{equation}
to construct

\begin{equation}
\begin{aligned}
%\begin{split}
\rho_{ij}\equiv & \mathds{P}(n_{i}, n_{j} 
\geq 1\vert \p,N) \\
= & 1+(1-p_i-p_j)^{N}
%\: & \hspace{3mm} -\sum_{n_i=1}^N {N \choose n_i} p_i^{n_i} (1-p_i-p_j)^{N-n_i} \\
%\: & \hspace{3mm} -\sum_{n_j=1}^N {N \choose n_j} p_j^{n_j} (1-p_i-p_j)^{N-n_j}\\
-(1-p_{i})^{N}-(1-p_{j})^{N}.
\label{rhoijN}
%\end{split}
\end{aligned}
\end{equation}
Higher moments can be straightforwardly computed using quantities such as 

\begin{equation}
\begin{aligned}
\rho_{ijk} & \equiv \mathds{P}(n_{i}, n_{j}, n_{k} \geq 1 
\vert \p, N) \\
\: & = 1 -(1-p_{i}-p_{j}-p_{k})^{N}\!-\!\!\sum_{\ell=i,j,k}\!\!(1-p_{\ell})^{N}
+\!\!\!\!\sum_{q\neq \ell=i,j,k}\!\!\!(1-p_{q}-p_{\ell})^{N}.
\end{aligned}
\end{equation}

These expressions arise when we compute the moments of $R$ [defined by
  Eq.~\eqref{eq:richness_one}] in terms of the probabilities $\p$.
Using the single-individual multinomial probability $\mathds{P}(\n
\vert \p,N)$ (Eq.~\eqref{MULTINOMIAL_n}) allows us to express moments
of the richness in a single individual in terms of the underlying
system probabilities $\p$. The first two are given by
\begin{equation}
\begin{aligned}
  \mathds{E}[R(\p)] & = \sum_{\n}\sum_{i=1}^{\Omega} \mathds{1}(n_{i}\geq 1)
  \mathds{P}(\n\vert \p, N) \\
\: & = \sum_{i=1}^{\Omega}\mathds{P}(n_{i}\geq 1\vert \p, N)=\sum_{i=1}^{\Omega}\rho_{i}, \\
\mathds{E}[R^{2}(\p)] & = \sum_{\n} \left[\sum_{i=1}^{\Omega}\mathds{1}(n_i\geq 1)\right]^{2}
\mathds{P}(\n\vert \p, N)\\
\: & = \sum_{i,j=1}^{\Omega}\mathds{P}(n_{i},n_{j} \geq 1\vert \p, N)
\equiv \mathds{E}[R]+ \sum_{j\neq i}^{\Omega}\rho_{ij}.
\label{R_MOMENT1}
\end{aligned}
\end{equation}
\subsection{Multi-individual quantities}
\label{sec:collective_quantities}
Here, we consider the distribution $\n^{(m)}$ across different
individuals and construct quantities describing group properties.  For
example, the combined richness of all TCR clones of $M$ individuals is
defined as

\begin{equation}
R^{(M)}(\{\n^{(m)}\}) \coloneqq
\sum_{k\geq 1}\sum_{i=1}^{\Omega}\mathds{1}\big(\textstyle{\sum\limits_{m=1}^{M}}n_{i}^{(m)}\!,
k\big).
\label{RICHNESS_M}
\end{equation}
To express the expected multi-individual richness in terms of the
underlying individual systems probabilities $\p^{(m)}$, we weight
Eq.~\eqref{RICHNESS_M} over the $M$-individual probability

\begin{equation}
\mathds{P}_{M}(\{\n^{(m)}\}\vert \{\p^{(m)}, N^{(m)}\})
\equiv 
\prod_{m=1}^{M}\mathds{P}(\n^{(m)}\vert \p^{(m)}, N^{(m)}),
\end{equation}
and sum over all allowable $\n^{(m)}$. For computing the first two
moments of the total-population richness in terms of $\p^{(m)}$, we
will make use of the marginalized probability $\tilde{\rho}_{i}$ that
clone $i$ appears in at least one of the $M$ individuals

\begin{equation}
\begin{aligned}
\tilde{\rho}_{i}\equiv & \mathds{P}\big(\textstyle{\sum\limits_{m=1}^{M}}n_{i}^{(m)}
\geq 1 \vert \{ \p^{(m)}, N^{(m)} \}\big) = 
1-\mathds{P}\big(n_{i}^{(m)}=0\,\, \forall\, m\big)\\
= & 1-\prod_{m=1}^{M}\big(1-p_{i}^{(m)}\big)^{N^{(m)}}.
\label{rhoiM}
\end{aligned}
\end{equation}
Note that $\tilde{\rho}_{i} >\prod_{m=1}^{M}\rho_{i}^{(m)}$ describes
the probability that a type $i$ cell occurs at all in the total
population, while $\prod_{m=1}^{M}\rho_{i}^{(m)}$ describes the
probability that a type $i$ cell appears in each of the $M$
individuals.

We will also need the joint probability that clones $i$ and $j$ both
appear in at least one of the $M$ individuals
$\mathds{P}\big(\sum_{m=1}^{M} n_{i}^{(m)} \geq 1,
\sum_{\ell=1}^{M}n_{j}^{(\ell)} \geq 1\vert \{\p^{(m)},
N^{(m)}\}\big)$, which we can decompose as

\begin{equation}
\begin{aligned}
& \mathds{P}\big(\textstyle{\sum\limits_{m=1}^{M}}n_{i}^{(m)}\geq 1,
\textstyle{\sum\limits_{\ell=1}^{M}}n_{j}^{(\ell)} \geq 1\vert \{\p^{(m)},N^{(m)}\}\big) \\
& \hspace{2.4cm} =  1-\mathds{P}\big(\textstyle{\sum\limits_{m=1}^{M}}n_{i}^{(m)} \geq 1,
\textstyle{\sum\limits_{\ell=1}^{M}}n_{j}^{(\ell)} =0 \vert \{\p^{(m)}, N^{(m)}\}\big) \\
& \hspace{3cm} - \mathds{P}\big(\textstyle{\sum\limits_{m=1}^{M}}n_{i}^{(m)} =0,
\textstyle{\sum\limits_{\ell=1}^{M}}n_{j}^{(\ell)} \geq 1 \vert \{\p^{(m)}, N^{(m)}\}\big) \\
& \hspace{3cm}- \mathds{P}\big(\textstyle{\sum\limits_{m=1}^{M}}n_{i}^{(m)} =
\textstyle{\sum\limits_{\ell=1}^{M}}n_{j}^{(\ell)}=0 \vert \{\p^{(m)}, N^{(m)}\}\big).
\label{probM}
\end{aligned}
\end{equation}
Upon using Eqs.~\eqref{MULTINOMIAL_n} and \eqref{ninj}, we find

\begin{equation}
\begin{aligned}
& \mathds{P}\big(\textstyle{\sum\limits_{m=1}^{M}}n_{i}^{(m)} \geq 1,
\textstyle{\sum\limits_{\ell=1}^{M}}n_{j}^{(\ell)} =0 \vert \{ \p^{(m)}, N^{(m)}\}\big)\\
& \hspace{3cm} = \prod_{m=1}^{M}\!\big(1-p_{i}^{(m)}\big)^{N^{(m)}}\!\!\!\!
- \prod_{m=1}^{M}\!\big(1-p_{i}^{(m)}\!\!\!-p_{j}^{(m)}\big)^{N^{(m)}},
\label{PP}
\end{aligned}
\end{equation}
allowing us to rewrite Eq.~\eqref{probM} as 

\begin{equation}
\begin{aligned}
\tilde{\rho}_{ij}\equiv & \mathds{P}\big(\textstyle{\sum\limits_{m=1}^{M}}n_{i}^{(m)}\geq 1,
\textstyle{\sum\limits_{\ell=1}^{M}}n_{j}^{(\ell)}\!\geq 1\vert \{\p^{(m)},
N^{(m)}\}\big)\\
= & 1 -\!\prod_{m=1}^{M}\!\big(1-p_{i}^{(m)}\big)^{N^{(m)}}\!\!\!\!-
\prod_{m=1}^{M}\!\big(1-p_{j}^{(m)}\big)^{N^{(m)}} \\[-2pt]
\: & \hspace{3cm} + \prod_{m=1}^{M}\!\big(1-p_{i}^{(m)}\!\!\!-p_{j}^{(m)}\big)^{N^{(m)}}\!\!\!\!\!.
\label{rhoijM}
\end{aligned}
\end{equation}
%
%are different
%from $\prod_{m=1}^{M}\rho_{i}^{(m)}$ and
%$\prod_{m=1}^{M}\rho_{ij}^{(m)}$,
%
%\begin{equation}
%\begin{aligned}
%  \tilde{\rho}_{i} & \equiv 1-\prod_{m=1}^{M}\big(1-p_{i}^{(m)}\big)^{N^{(m)}}
%  \not= \prod_{m=1}^{M}\rho_{i}^{(m)}\equiv 
% \not=  \prod_{m=1}^{M}\left[1-(1-p_{i}^{(m)})^{N^{(m)}}\right]\\
%  \tilde{\rho}_{ij} & \equiv 1 - \prod_{m=1}^{M}\!\big(1-p_{i}^{(m)}\big)^{N^{(m)}}\!\!\!\!\!
%  - \!\prod_{m=1}^{M}\!\big(1-p_{j}^{(m)}\big)^{N^{(m)}}\!\!\!\!\!+
%\prod_{m=1}^{M}\!\big(1-p_{i}^{(m)}-p_{j}^{(m)}\big)^{N^{(m)}} \\
%\: &  \not= \!\prod_{m=1}^{M}\bigg[1-\big(1-p_{i}^{(m)}\big)^{N^{(m)}}\!\!\!\!\!\!
%-\big(1-p_{i}^{(m)}\big)^{N^{(m)}}\!\!\!\!\!+ \big(1-p_{i}^{(m)}-p_{j}^{(m)}\big)^{N^{(m)}}\bigg]
% \\ \: & \not= \!\prod_{m=1}^{M}\!\rho_{ij}^{(m)},
%\label{NOTEQUAL}
%\end{aligned}
%\end{equation}
%
Note also that $\tilde{\rho}_{ij} > \prod_{m=1}^{M}\rho_{ij}^{(m)}$. 

Using the above definitions, we can express the mean total-population
richness as

\begin{equation}
\begin{aligned}
  \mathds{E}[R^{(M)}(\{\p^{(m)}\})] & =  \sum_{\n^{(m)}}\sum_{i=1}^{\Omega}
 \sum_{k\geq 1}\mathds{1}\big(\textstyle{\sum\limits_{\ell=1}^{M}}n_{i}^{(\ell)}\!,k\big)
\prod_{m=1}^{M}\mathds{P}(\n^{(m)}\vert \p^{(m)}, N^{(m)}) \\
\: & = \sum_{i=1}^{\Omega} \mathds{P}\big(\textstyle{\sum\limits_{m=1}^{M}}n_{i}^{(m)}
\geq 1\vert \{\p^{(m)}, N^{(m)}\}\big)\\
\: & = \sum_{i=1}^{\Omega}\Big[1-\prod_{m=1}^{M}(1-p_{i}^{(m)})^{N^{(m)}}\Big]\equiv 
\sum_{i=1}^{\Omega} \tilde{\rho}_{i} \\
\: & = \Omega - \sum_{i=1}^{\Omega}\prod_{m=1}^{M}(1-p_{i}^{(m)})^{N^{(m)}} \\
\: & \approx \Omega - \sum_{i=1}^{\Omega} e^{-\sum_{m=1}^{M}p_{i}^{(m)}N^{(m)}}\!\!\!, 
%\,\,\, p_{i}^{(m)}\ll 1,\, N^{(m)}\gg 1.
\label{RM_AVE}
\end{aligned}
\end{equation}
where the last approximation holds for $p_{i}^{(m)}\ll 1, N^{(m)}\gg
1$. The second moment of the total $M$-population richness can also be
found in terms of $\mathds{E}[R^{(M)}]$ and Eq.~\eqref{rhoijM}, 

\begin{align}
  \begin{split}
%\begin{equation}
%\begin{aligned}
  \mathds{E}\big[\big(R^{(M)}(\{\p^{(m)}\})\big)^{2}\big] = & \sum_{\{\n^{(M)}\}}\left[
    \sum_{i=1}^{\Omega}\mathds{1}\big(\textstyle{\sum\limits_{m=1}^{M}}n_{i}^{(m)}\geq 1\big)\right]^{2}
  \prod_{m=1}^{M}\mathds{P}\big(\n^{(m)} \vert \p^{(m)}, N^{(m)}\big)\\
= & \sum_{i,j=1}^{\Omega}\mathds{P}\big(\textstyle{\sum\limits_{m=1}^{M}}n_{i}^{(m)}\geq 1,
\textstyle{\sum\limits_{\ell=1}^{M}}n_{j}^{(\ell)}\geq 1 \vert \{\p^{(m)}, N^{(m)}\}\big) \\
=  &  \sum_{i=1}^{\Omega}\mathds{P}\big(\textstyle{\sum\limits_{m=1}^{M}}n_{i}^{(m)}\geq 1  
\vert \p^{(m)}, N^{(m)}\big) \\[-3pt]
\: & \qquad + \sum_{i\neq j}^{\Omega} 
\mathds{P}\big(\textstyle{\sum\limits_{m=1}^{M}}n_{i}^{(m)}\geq 1, 
\textstyle{\sum\limits_{\ell=1}^{M}}n_{j}^{(\ell)}\geq 1 \vert \{\p^{(m)}, N^{(m)}\}\big) \\
= &\, \mathds{E}[R^{(M)}(\{\p^{(m)}\})] + \sum_{i\neq j}^{\Omega}\tilde{\rho}_{ij}.
%
%\Omega^{2} -(2\Omega-1)\sum_{i=1}^{\Omega}
%\prod_{m=1}^{M}\big(1-p_{i}^{(m)}\big)^{N^{(m)}}
%+\sum_{i\neq j}^{\Omega}\prod_{m=1}^{M}\big(1-p_{i}^{(m)}-p_{j}^{(m)}\big)^{N^{(m)}}.
\label{RM_2}
%\end{aligned}
%\end{equation}
  \end{split}
  \end{align}

\begin{figure}
    \centering
    \includegraphics[width=0.96\textwidth]{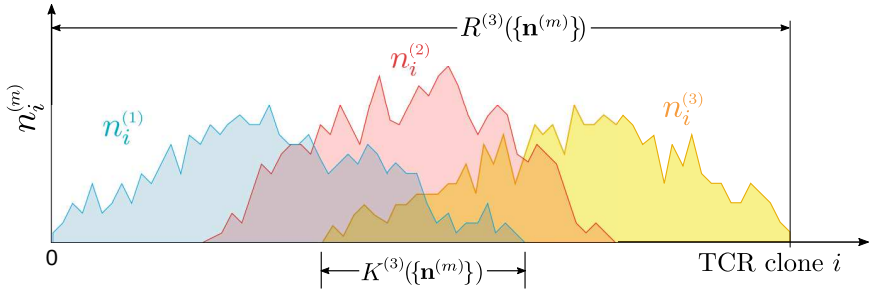}
    \caption{Three individuals with overlapping cell number
      distributions $n_{i}^{(m)}, m=1,2,3$. The richness $R^{(3)}$ is
      the total number of distinct number of different TCRs found
      across all individuals, and is defined in
      Eq.~\eqref{RICHNESS_M}. The overlap $K^{(3)}$ is the number of TCR
      clones found in all three individuals, as defined in
      Eq.~\eqref{KM}. For visual simplicity, the set of clones $i$
      present in each individual are drawn to be contiguous. When
      considering subsampling of cells from each individual, $K^{(M)}$
      will be reduced since some cell types $i$ will be lost. The
      corresponding values, $s_{i}^{(m)}$, $K_{\rm s}^{(M)}$, and
      $R_{\rm s}^{(M)}$ can be constructed from Eqs.~\eqref{sample_ns}
      and \eqref{sigma} reflecting the losses from subsampling.}
    \label{OVERLAP}
\end{figure}

Given $\n^{(m)}$ of all individuals, we can also easily define the
number of distinct TCR clones that appear in all of $M$ randomly
selected individuals, the ``$M$-overlap'' or ``$M$-publicness''

\begin{equation}
K^{(M)}(\{\n^{(m)}\})\coloneqq \sum_{i=1}^{\Omega}\prod_{m=1}^{M}\sum_{k^{(m)}\geq 1}
\!\mathds{1}(n_{i}^{(m)}\!,k^{(m)}).
\label{KM}
\end{equation}
Figure \ref{OVERLAP} provides a simple example of three individuals
each with a contiguous distribution of cell numbers $n_{i}^{(m)}$ that
overlap.

As with Eqs.~\eqref{eq:richness_one} and \eqref{R_MOMENT1}, we can
express the overlap in terms of the underlying individual
probabilities $\p^{(m)}$ by weighting Eq.~\eqref{KM} by the
$M$-population probability $\prod_{m=1}^{M} \mathds{P}(\n^{(m)}\vert
\p^{(m)},N^{(m)})$ (see Eq.~\eqref{MULTINOMIAL_n}) to find
\begin{equation}
\begin{aligned}
  \mathds{E}[K^{(M)}(\{\p^{(m)}\})] = & 
\sum_{\{\n^{(m)}\}}\left[\sum_{i=1}^{\Omega}\prod_{m=1}^{M} \mathds{1}(n_{i}^{(m)}\geq 1)\right]
\mathds{P}_{M}(\{\n^{(m)}\}\vert \{\p^{(m)}, N^{(m)}\})\\
= & \sum_{i=1}^{\Omega}\prod_{m=1}^{M}\mathds{P}(n_{i}^{(m)}\geq 1\vert \{\p^{(m)}, N^{(m)}\})\\
= & \sum_{i=1}^{\Omega}\prod_{m=1}^{M}\left[1-\big(1-p_{i}^{(m)}\big)^{N^{(m)}}\right]
\equiv \sum_{i=1}^{\Omega}\prod_{m=1}^{M}\rho_{i}^{(m)}
%
%\equiv  & \sum_{i=1}^{\Omega}\prod_{m=1}^{M}\rho_{i}^{(m)}
\label{K_M_p}
\end{aligned}
\end{equation}
\begin{equation}
\begin{aligned}
\: & \mathds{E}\big[\big(K^{(M)}(\{\p^{(m)}\})\big)^{2}\big]\\
\:  & \hspace{7mm} =  \sum_{\{\n^{(m)}\}} 
\left[\sum_{i=1}^{\Omega}\prod_{m=1}^{M} \mathds{1}(n_{i}^{(m)}\geq 1)\right]^{2}
 \mathds{P}_{M}(\{\n^{(m)}\}\vert \{\p^{(m)}, N^{(m)}\}) \\
\: &  \hspace{7mm} = \sum_{i,j=1}^{\Omega}\prod_{m=1}^{M}
 \mathds{P}(n_{i}^{(m)},n_{j}^{(m)}\geq 1\vert \{\p^{(m)}, N^{(m)}\})\\
\: & \hspace{7mm} = \sum_{i=1}^{\Omega}\prod_{m=1}^{M}
\left[1-(1-p_{i}^{(m)})^{N^{(m)}}\right] \\[-2pt]
\: & \hspace{9mm}+\sum_{j\neq i}^{\Omega}\prod_{m=1}^{M}\!
\Big[1+(1-p_{i}^{(m)}\!\!-p_{j}^{(m)})^{N^{(m)}}-(1-p_{i}^{(m)})^{N^{(m)}}\!\!\!\!-(1-p_{j}^{(m)})^{N^{(m)}}\Big]\\[3pt]
\: &  \hspace{7mm} \equiv \mathds{E}[K^{(M)}(\{\p^{(m)}\})]+\sum_{j\neq i}^{\Omega}\prod_{m=1}^{M}\rho_{ij}^{(m)}.
\label{K_M_p_2}
\end{aligned}
\end{equation}
The expected number of clones shared among all $M$ individuals,
$\mathds{E}[K^{(M)}]$, provides a natural measure of
$M$-overlap. Clearly, $\mathds{E}[K^{(M)}]<\mathds{E}[K^{(M')}]$ if
$M>M'$. As with $M$-publicness, we can identify \textbf{the expected
  $M$-privateness} as
$\Omega-\mathds{E}\big[K^{(M)}\big]$, the expected number of clones
that are not shared by all $M$ individuals, \textit{i.e.}, that occur
in at most $M-1$ individuals. This ``privateness'' is related to a
multi-distribution generalization of the ``dissimilarity probability''
of samples from two different discrete distributions
\cite{hampton2012estimation}. Variations in $M$-overlap associated
with a certain cell-type distribution are captured by the variance
${\rm
  Var}\big[K^{(M)}\big]=\mathds{E}\big[(K^{(M)})^{2}\big]-\mathds{E}\big[K^{(M)}\big]^2$.
If the total number of sequences $\Omega$ is very large,
parallelization techniques (see Sec.~\ref{sec:simus}) should be
employed to evaluate the term $\sum_{j\neq
  i}^{\Omega}\prod_{m=1}^{M}\rho_{ij}^{(m)}$ in
$\mathds{E}\big[(K^{(M)})^{2}\big]$.

A more specific definition of overlap or privateness may be that a
clone must appear in at least some specified fraction of $M$ tested
individuals.  To find the probability that $M_{i} \leq M$ individuals
share at least one cell of a single type $i$, we use the Poisson
binomial distribution describing independent Bernoulli trials on
individuals with different success probabilities $\rho_{i}^{(m)}
\equiv \rho(n_{i}^{(m)} \geq 1)$:

\begin{equation}
\mathds{P}(M_{i}\vert \{p_{i}^{(m)}\}) = \sum_{A\in F_{M_{i}}}\prod_{m \in A}\rho_{i}^{(m)}
\prod_{\ell\in A^{\rm c}}(1-\rho_{i}^{(\ell)}),
\label{PmN}
\end{equation}
where $F_{M_{i}}$ is the set of all subsets of $M_{i}$ integers that
can be selected from the set $(1,2,3,\ldots,M)$ and $A^{\rm c}$ is the
complement of $A$. Equation~\eqref{PmN} gives a probabilistic measure
of the prevalence of TCR $i$ across $M$ individuals. For example, one
can use it to define a mean frequency $\mathds{E}[M_{i}]/M$. One can
evaluate Eq.~\eqref{PmN} recursively or using Fourier transforms,
particularly for $M < 20$ \cite{CHEN_1997,HONG_2013}.

\subsection{Subsampling} 
\label{sec:subsampling}
The results above are described in terms of the entire cell
populations $\n^{(m)}$ or their intrinsic generation probabilities
$\p^{(m)}$.  In practice, one cannot measure $n_{i}^{(m)}$ or
even $N^{(m)}$ in any individual $m$. Rather, we can only sample a
much smaller number of cells $S^{(m)} \ll N^{(m)}$ from individual
$m$, as shown in Fig.~\ref{FIG2}. Within this subsample from
individual $m$, we can count the number $s_{i}^{(m)}$ of type-$i$
cells. Since only subsamples are available, we wish to define
quantities such as probability of occurrence, richness, and overlap in
terms of the cell counts $\s^{(m)} \equiv \{s_{i}^{(m)}\}$ in the
sample extracted from an individual.
%
%Since the quantities in Eqs.~\eqref{R_MOMENTS}, \eqref{RM_AVE}, and
%\eqref{K_M_p} are ``intensive'' the same functional forms are used to
%define the corresponding properties of the subsamples by simply
%replacing $\n$ with $\s$ and $N^{(m)}$ with $S^{(m)}$.
%
Quantities such as \textit{sampled} richness and overlap can be
defined in the same way except with $\s^{(m)}$ as the underlying
population configuration.  To start, first assume that the cell count
$\n$ in a specific individual is given. If that individual has $N$
cells of which $S$ are sampled, the probability of observing the
population $\s = \{s_{1}, s_{2}, \ldots, s_{\Omega}\}$ in the sample
is given by (assuming all cells are uniformly distributed and randomly
subsampled at once, without replacement) \cite{CHAO}

\begin{equation}
\begin{aligned}
\mathds{P}(\s\vert \n, S, N) & = {1\over {N \choose S}}
\prod_{i=1}^{\Omega}{n_{i} \choose s_{i}},\quad \sum_{i=1}^{\Omega}s_{i} = S.
%\mathds{1}(S,\sum_{j=1}^{\Omega} s_{j}).
\label{sample_ns}
\end{aligned}
\end{equation}
The probability that cell type $j$ appears in the sample from an
individual with population $\n$ can be found by marginalizing over all
$s_{j\neq i}$, giving

\begin{equation}
\begin{aligned}
\sigma_{i} \equiv \mathds{P}(s_{i}\geq 1 \vert \n, S, N) & \displaystyle =
1-{{N-n_{i}\choose S}\over {N\choose S}}.
\label{sigma}
\end{aligned}
\end{equation}
This result can be generalized to more than one TCR clone present. For
example, the probability that both clones $i$ and $j$ are found in a
sample is

\begin{equation}
\begin{aligned}
\sigma_{ij}\equiv  \mathds{P}(s_{i}, s_{j} \geq 1 \vert \n, S, N) & = \displaystyle
  1+{{N-n_{i}-n_{j}\choose S}\over {N\choose S}}-
  {{N-n_{i}\choose S}\over {N\choose S}}-{{N-n_{j}\choose S}\over {N\choose S}}.
  \label{sigmaij}
\end{aligned}
\end{equation}

Using Eq.~\eqref{sample_ns} as the probability distribution,
we can also find the probability that clone $i$
appears in any of the $M$ $S^{(m)}$-sized samples

\begin{equation}
\begin{aligned}
  \tilde{\sigma}_{i} \equiv \mathds{P}\big(\textstyle{\sum\limits_{m=1}^{M}}s_{i}^{(m)}\geq 1
  \vert \{\n^{(m)}, S^{(m)}, N^{(m)}\}\big) & \displaystyle =
1-\prod_{m=1}^{M}{{N^{(m)}-n_{i}^{(m)}\choose S^{(m)}}\over {N^{(m)}\choose S^{(m)}}},
\label{sigmaiM}
\end{aligned}
\end{equation}
and the joint probabilities that clones $i$ and $j$ appear in any sample

\begin{equation}
\begin{aligned}
  \tilde{\sigma}_{ij}\equiv & \mathds{P}\big(\textstyle{\sum\limits_{m=1}^{M}}s_{i}^{(m)}\geq 1,
  \textstyle{\sum\limits_{\ell=1}^{M}}s_{j}^{(\ell)} \geq 1
  \vert \{\n^{(m)}, S^{(m)}, N^{(m)}\}\big) \\[4pt]
  = & \displaystyle
  1+\prod_{m=1}^{M}\!{{N^{(m)}-n_{i}^{(m)}-n_{j}^{(m)}\choose S^{(m)}}\over {N^{(m)}\choose S^{(m)}}}-
  \prod_{m=1}^{M}\!{{N^{(m)}-n_{i}^{(m)}\choose S^{(m)}}\over {N^{(m)}\choose S^{(m)}}}
  -\prod_{m=1}^{M}\!{{N^{(m)}-n_{j}^{(m)}\choose S^{(m)}}
    \over {N^{(m)}\choose S^{(m)}}}.
\label{sigmaijM}
\end{aligned}
\end{equation}

Quantities such as richness and publicness \textit{measured within
  samples} from the group can be analogously defined in terms of
clonal populations $\s^{(m)}$:

\begin{equation}
R_{\rm s}(\s)\coloneqq  \sum_{k\geq 1}\sum_{i=1}^{\Omega} \mathds{1}(s_{i}, k),
\label{RICHNESS_S}
\end{equation}

\begin{equation}
R_{\rm s}^{(M)}(\{\s^{(m)}\}) \coloneqq
\sum_{k\geq1}\sum_{i=1}^{\Omega}\mathds{1}\big(\textstyle{\sum\limits_{m=1}^{M}}s_{i}^{(m)},
k\big),
\label{RICHNESS_S_M}
\end{equation}
and

\begin{equation}
K_{\rm s}^{(M)}(\{\s^{(m)}\})\coloneqq \sum_{i=1}^{\Omega}\prod_{m=1}^{M} \sum_{k^{(m)}\geq 1}
\mathds{1}(s_{i}^{(m)},k^{(m)}).
\label{K_S_M}
\end{equation}
For a given $\n^{(m)}$, these quantities can be first averaged over
the sampling distribution Eq.~\eqref{sample_ns} to express them in
terms of $\n^{(m)}$ and to explicitly reveal the effects of random
sampling.  The first two moments of $R_{\rm s}$, $R^{(M)}_{\rm s}$,
and $K_{\rm s}^{(M)}$ expressed in terms of $\n^{(m)}$ can be easily
found by weighting Eqs.~\eqref{RICHNESS_S}, ~\eqref{RICHNESS_S_M}, and
\eqref{K_S_M} by $\mathds{P}(\s\vert \n, S, N)$ and $P^{(M)} =
\prod_{m=1}^{M}\mathds{P}(\s^{(m)} \vert \n^{(m)}, S^{(m)}, N^{(m)})$:

\begin{equation}
  \begin{aligned}
    \mathds{E}[R_{\rm s}(\n)] = & \,\Omega -
        {1\over {N \choose S}}\sum_{i=1}^{\Omega} {N-n_{i}\choose S}\equiv 
\sum_{i=1}^{\Omega} \sigma_{i},\\
        \mathds{E}\big[\big(R_{\rm s}(\n)\big)^{2}\big] = & \,
        \mathds{E}[R_{\rm s}(\n)]+\sum_{i\neq j}^{\Omega}\sigma_{ij}
\label{R_n}
\end{aligned}
  \end{equation}

\begin{equation}
\begin{aligned}
%\begin{split}
  \: & \mathds{E}\big[R_{\rm s}^{(M)}(\{\n^{(m)}\})\big]\\[-5pt]
  \: & \hspace{6mm} =\sum_{\s^{(m)}}\sum_{i=1}^{\Omega}
\mathds{1}(\textstyle{\sum\limits_{m=1}^M} s_{i}^{(m)}\geq 1)
\prod_{m=1}^{M}\mathds{P}(\s^{(m)} \vert \n^{(m)}\!, S^{(m)}\!, N^{(m)})\\
\: & \hspace{6mm} = \sum_{i=1}^{\Omega}\mathds{P}(\textstyle{\sum\limits_{m=1}^M} s_{i}^{(m)}\geq 1
\vert \{\n^{(m)}\!, S^{(m)}\!, N^{(m)}\})\\
\: & \hspace{6mm} = \Omega-\sum_{i=1}^\Omega \prod_{m=1}^M {{N-n_{i}^{(m)}\choose S}\over {N\choose S}}
\equiv \sum_{i=1}^\Omega\tilde{\sigma}_{i}, \\[4pt]
%
%\: &=\Omega-\sum_{i=1}^{\Omega}
%\mathds{P}(\textstyle{\sum\limits_{m=1}^M} s_{i}^{(m)}=0 \vert \{\n^{(m)}, S^{(m)}, N^{(m)}\})\\
%\: & =\Omega-\sum_{i=1}^\Omega \prod_{m=1}^M {{N-n_{i}^{(m)}\choose S}\over {N\choose S}} \\
%\: & \equiv \sum_{i=1}^\Omega\tilde{\sigma}_{i}, \\
%
\: & \mathds{E}\Big[\big(R_{\rm s}^{(M)}(\{\n^{(m)}\})\big)^{2}\Big]\\[-5pt]
\: & \hspace{8mm} =\sum_{\n^{(m)}}\left[\sum_{i=1}^{\Omega}
\mathds{1}(\textstyle{\sum\limits_{m=1}^M} s_{i}^{(m)}\geq 1)\right]^{2}
\prod_{m=1}^{M}\mathds{P}(\s^{(m)} \vert \n^{(m)}, S^{(m)}, N^{(m)}) \\
\: & \hspace{8mm}= \sum_{i,j=1}^{\Omega}\prod_{m=1}^{M}\mathds{P}
\big(\textstyle{\sum\limits_{m=1}^M} s_{i}^{(m)},
\textstyle{\sum\limits_{\ell=1}^M} s_{j}^{(m)}\geq 1 \vert \{\n^{(m)}, S^{(m)}, N^{(m)}\})\\[-3pt]
\: & \hspace{8mm} = \mathds{E}\big[R_{\rm s}^{(M)}(\{\n^{(m)}\})\big]
+ \!\sum_{i\neq j=1}^\Omega\!\tilde{\sigma}_{ij},
\label{RICHNESS_S_M_n}
%\end{split}
\end{aligned}
\end{equation}
\vspace{-2mm}
\begin{equation}
\begin{aligned}
\hspace{3mm} \mathds{E}\big[K_{\rm s}^{(M)}(\{\n^{(m)}\})\big] & \: \\[-3pt]
\: & \hspace{-20mm} =\! \sum_{\{\s^{(m)}\}}
\sum_{i=1}^{\Omega}\prod_{m=1}^{M}\!\mathds{1}(s_{i}^{(m)}\geq 1)
\mathds{P}(\s^{(m)} \vert \n^{(m)}\!, S^{(m)}\!, N^{(m)})\\
\: &   \hspace{-20mm}  = \sum_{i=1}^{\Omega}\prod_{m=1}^{M}
\!\mathds{P}(s_{i}^{(m)}\geq 1 \vert \{\n^{(m)}\!, S^{(m)}\!, N^{(m)}\})\\
%
%\: & = \sum_{i=1}^{\Omega} \prod_{m=1}^{M}{1\over {N^{(m)}\choose S^{(m)}}}\sum_{k^{(m)}\geq 1}
%\int_{0}^{2\pi}{\dd q^{(m)}\over 2\pi}(1+e^{iq^{(m)}})^{N^{(m)}-n_{i}^{(m)}} \\[-6pt]
%\: & \hspace{4.8cm} \times {n_{i}^{(m)}\choose k^{(m)}}e^{-iq^{(m)}S^{(m)}}e^{iq^{(m)}k^{(m)}} \\
%\: & =  \sum_{i=1}^{\Omega} \prod_{m=1}^{M}{1\over {N^{(m)}\choose S^{(m)}}}   
%\sum_{k^{(m)}\geq 1}{N^{(m)}-n_{i}^{(m)} \choose S^{(m)}-k^{(m)}}{n_{i}^{(m)}\choose k^{(m)}} \\
%
\: &  \hspace{-20mm} = \sum_{i=1}^{\Omega} \prod_{m=1}^{M}\!\left[1-\frac{{N^{(m)}-n_{i}^{(m)}\choose S^{(m)}}}
  {{N^{(m)}\choose S^{(m)}}}\right]  \equiv \sum_{i=1}^{\Omega} \prod_{m=1}^{M}\!\sigma_{i}^{(m)},\\[4pt]
%\nonumber
%\end{aligned}
%\end{equation}
%
%\begin{eqnarray}
%\begin{aligned}
%
\mathds{E}\Big[\big(K_{\rm s}^{(M)}(\{\n^{(m)}\})\big)^{2}\Big] & \: \\[-1pt]
\:  & \hspace{-20mm} = \sum_{\{\s^{(m)}\}}\left[\sum_{i=1}^{\Omega} \prod_{m=1}^{M}
  \mathds{1}(s_{i}^{(m)}\geq 1)\right]^{2}\prod_{m=1}^{M}
\mathds{P}(\s^{(m)} \vert \n^{(m)}\!\!, S^{(m)}\!\!, N^{(m)})   \\
\:  & \hspace{-20mm}  = \sum_{i,j=1}^{\Omega}\prod_{m=1}^{M}
\mathds{P}(s_{i}^{(m)},s_{j}^{(m)}\geq 1 \vert \{\n^{(m)}\!\!, S^{(m)}\!\!, N^{(m)}\}) \\[-3pt]
\:  & \hspace{-20mm} \equiv \mathds{E}\big[K_{\rm s}^{(M)}(\{\n^{(m)}\})\big]
+\sum_{i\neq j}^{\Omega} \prod_{m=1}^{M}\sigma_{ij}^{(m)}.
\label{K_S_M_n}
\end{aligned}
\end{equation}

All of the above quantities can also be expressed in terms of the
underlying probabilities $\p^{(m)}$ rather than the population
configurations $\n^{(m)}$. To do so, we can further weight
Eqs.~\eqref{R_n}, \eqref{RICHNESS_S_M_n}, and \eqref{K_S_M_n} over the
probability Eq.~\eqref{MULTINOMIAL_n} to render these quantities in
terms of the underlying probabilities $\p^{(m)}$. However, we
can also first convolve Eq.~\eqref{sample_ns} with the multinomial
distribution in Eq.~\eqref{MULTINOMIAL_n} (suppressing the individual
index $m$)

\begin{equation}
\begin{aligned}
\mathds{P}(\s \vert \p, S, N) & = \sum_{\n} \mathds{P}(\s \vert \n, S, N)
\mathds{P}(\n \vert \p, N),
\label{CONVOL}
\end{aligned}
\end{equation}
along with the implicit constraints $\sum_{i=1}^{\Omega}n_{i} \equiv N$ and
$\sum_{i=1}^{\Omega}s_{i} = S$ to find 

\begin{equation}
\begin{aligned}
\mathds{P}(\s\vert \p, S) & = S! \prod_{i=1}^{\Omega} 
{p_{i}^{s_{i}}\over s_{i}!},\quad \sum_{i=1}^{\Omega}s_{i} = S,
\label{Psp}
\end{aligned}
\end{equation}
which is a multinomial distribution identical in form to $\mathds{P}(\n
\vert \p, N)$ in Eq.~\eqref{MULTINOMIAL_n}, except with $\n$ replaced
by $\s$ and $N$ replaced by $S$. Uniform random sampling from a
multinomial results in another multinomial.  Thus, if we use the full
multi-individual probability
\begin{equation}
  \mathds{P}_{M}(\{\s^{(m)}\} \vert \{\p^{(m)}\!,S^{(m)}\})
\equiv \prod_{m=1}^{M}\mathds{P}(\s^{(m)} \vert \p^{(m)}\!, S^{(m)})
\end{equation}
to compute moments of the sampled richness and publicness, they take
on the same forms as the expressions associated with the
whole-organism quantities.
For example, in the $\p$ representation, the probability that clone
$i$ appears in the sample from individual $m$ is

\begin{equation}
  \rho_{i}^{(m)}(S)\equiv \mathds{P}(s_{i}^{(m)}\geq 1 \vert \{\p^{(m)}\!, S^{(m)}\})
  = 1-(1-p_{i}^{(m)})^{S^{(m)}}\!\!,
\label{rhoS}
\end{equation}
in analogy with Eq.~\eqref{rhoN}, while the two-clone joint probability
in the sampled from individual $m$ becomes

\begin{equation}
\begin{aligned}
\rho_{ij}^{(m)}(S)\equiv & \mathds{P}(s^{(m)}_{i}, s^{(m)}_{j} 
\geq 1\vert \{\p^{(m)}\!,S^{(m)}\})\\
= & 1+(1-p^{(m)}_{i}\!\!-p_{j}^{(m)})^{S^{(m)}}\!\!\!
-(1-p_{i}^{(m)})^{S^{(m)}}\!\!\!-(1-p_{j}^{(m)})^{S^{(m)}},
\label{rhoijS}
\end{aligned}
\end{equation}
in analogy with Eq.~\eqref{rhoijN}. Similarly, for the overlap quantities,
in analogy with Eqs.~\eqref{rhoiM} and
\eqref{rhoijM}, we have 

\begin{equation}
\begin{aligned}
\tilde{\rho}_{i}(S)\equiv & \mathds{P}\big(\textstyle{\sum\limits_{m=1}^{M}}s_{i}^{(m)}
\geq 1 \vert \{ \p^{(m)}, S^{(m)} \}\big) = 
1-\mathds{P}\big(s_{i}^{(m)}=0\,\, \forall\, m\big)\\
= & 1-\prod_{m=1}^{M}\big(1-p_{i}^{(m)}\big)^{S^{(m)}}.
\label{rhoiMS}
\end{aligned}
\end{equation}
\begin{equation}
\begin{aligned}
\tilde{\rho}_{ij}(S)\equiv & \mathds{P}\big(\textstyle{\sum\limits_{m=1}^{M}}s_{i}^{(m)}\geq 1,
\textstyle{\sum\limits_{\ell=1}^{M}}s_{j}^{(\ell)} \geq 1\vert \{\p^{(m)},
S^{(m)}\}\big) \\
= & 1-\!\!\prod_{m=1}^{M}\!\big(1-p_{i}^{(m)}\big)^{S^{(m)}}\!\!\!\!-\!\prod_{m=1}^{M}\!
\big(1-p_{j}^{(m)}\big)^{S^{(m)}}
\!\!\!\!+\!\prod_{m=1}^{M}\!\big(1-p_{i}^{(m)}\!\!\!-p_{j}^{(m)}\big)^{S^{(m)}}\!\!.
\label{rhoijMS}
\end{aligned}
\end{equation}
The expressions for the sampled moments $\mathds{E}[R_{\rm s}(\p)]$,
$\mathds{E}[R_{\rm s}^{2}(\p)]$, $\mathds{E}\big[R_{\rm
    s}^{(M)}(\{\p^{(m)}\})\big]$, $\mathds{E}\big[\big(R_{\rm
    s}^{(M)}(\{\p^{(m)}\})\big)^{2}\big]$, $\mathds{E}[K_{\rm
    s}^{(M)}(\{\p^{(m)}\})]$, and $\mathds{E}\big[\big(K_{\rm
    s}^{(M)}(\{\p^{(m)}\})\big)^{2}\big]$ follow the same form as
their unsampled counterparts given in Eqs.~\eqref{R_MOMENT1},
\eqref{RM_AVE}, \eqref{RM_2}, \eqref{K_M_p}, and \eqref{K_M_p_2},
except with $\rho_{i}^{(m)}$, $\rho_{ij}^{(m)}$, $\tilde{\rho}_{i}$,
and $\tilde{\rho}_{ij}$ replaced by their $\rho_{i}^{(m)}(S)$,
$\rho_{ij}^{(m)}(S)$, $\tilde{\rho}_{i}(S)$, and
$\tilde{\rho}_{ij}(S)$ counterparts. This simplifying property
  arises because of the conjugate nature of the multinomial
  distributions \eqref{MULTINOMIAL_n}, \eqref{Psp}, and
  \eqref{CONVOL}.

In addition to simple expressions for the moments of $K_{\rm
  s}^{(M)}$, we can also find expressions for the probability
distribution over the values of $K_{\rm s}^{(M)}$.  In terms of
$\n^{(m)}$, since the probability that $s_{i}^{(m)}\geq 1$ in the
samples from all $1\leq m \leq M$ individuals is $\sigma_{i} \equiv
\prod_{m=1}^{M} \mathds{P}(s_{i}^{(m)}\geq 1\vert \{\n^{(m)}, S^{(m)},
N^{(m)}\}) = \prod_{m=1}^{M}\sigma_{i}^{(m)}$, the probability that
exactly $k$ clones are shared by all $M$ samples is

\begin{equation}
\mathds{P}(K^{(M)}_{\rm s} = k \vert \{\sigma_{i}^{(m)}\}) 
= \sum_{A\in F_{k}}\prod_{i \in A}\left(\prod_{m=1}^{M}\sigma_{i}^{(m)}\right)
\prod_{j\in A^{\rm c}}\left[1-\prod_{m=1}^{M}\sigma_{j}^{(m)}\right],
\label{PK=k}
\end{equation}
where $F_{k}$ is the set of all subsets of $k$ integers that can be
selected from the set $\{1,2,3,\ldots, K^{(M)}\}$ and $A^{\rm c}$ is
the complement of $A$. Equation~\eqref{PK=k} is the Poisson
binomial distribution, but this time the underlying success
probabilities $\prod_{m=1}^{M}\sigma_{i}^{(m)}$ across all $M$
individuals vary with TCR clone identity $i$.

Finally, inference of individual measures from subsamples can be
formulated.  One can use the sampling likelihood function
$\mathds{P}(\s \vert \n, S, N)$, Bayes' rule, and the multinomial
(conjugate) prior $\mathds{P}(\n \vert \p, N)$ to construct the
posterior probability of $\n$ \textit{given} a sampled configuration
$\s$:

\begin{equation}
\begin{aligned}
  \mathds{P}(\n \vert \s, S, N, \p) & = \frac{\mathds{P}(\s \vert \n, S, N)
    \mathds{P}(\n \vert \p, N)}{\sum_{\n} \mathds{P}(\s \vert \n, S, N)
    \mathds{P}(\n \vert \p, N)}.
\label{BAYES}
\end{aligned}
\end{equation}
%
%We then use Eq.~\eqref{sample_ns} for the likelihood $\mathds{P}(\s
%\vert \n, S, N)$ and Eq.~\eqref{MULTINOMIAL_n} for the conjugate prior
%$\mathds{P}(\n \vert \p, N)$ (where $\p$ are hyperparameters).  
%
The normalization in Eq.~\eqref{BAYES} has already been found in
Eqs.~\eqref{CONVOL} and \eqref{Psp}.  Thus, we find the posterior

\begin{equation}
  \mathds{P}(\n \vert \s, S, N, \p) =
  (N-S)! \prod_{i=1}^{\Omega}\frac{p_{i}^{n_{i}-s_{i}}}{(n_{i}-s_{i})!},\quad
  \sum_{i=1}^{\Omega}(n_{i}-s_{i}) = N-S
\end{equation}
in terms of the hyperparameters $\p$.  Using this posterior, we can
calculate the expectation of the whole organism richness
$R=\sum_{k\geq 1}\sum_{i=1}^{\Omega} \mathds{1}(n_{i}, k)$,

\begin{equation}
  \mathds{E}[R(\s, \p)] = \Omega - \sum_{j\vert
    s_{j}=0}\left(1-p_{j}\right)^{N-S},
\label{INFER_R}
\end{equation}
which depends on the sampled configuration only through the
sample-absent clones $j$. Bayesian methods for estimating overlap
between two populations from samples have also been recently explored
\cite{larremore2019}.

\begin{figure}[htb]
\centering
\includegraphics[width=4.7in]{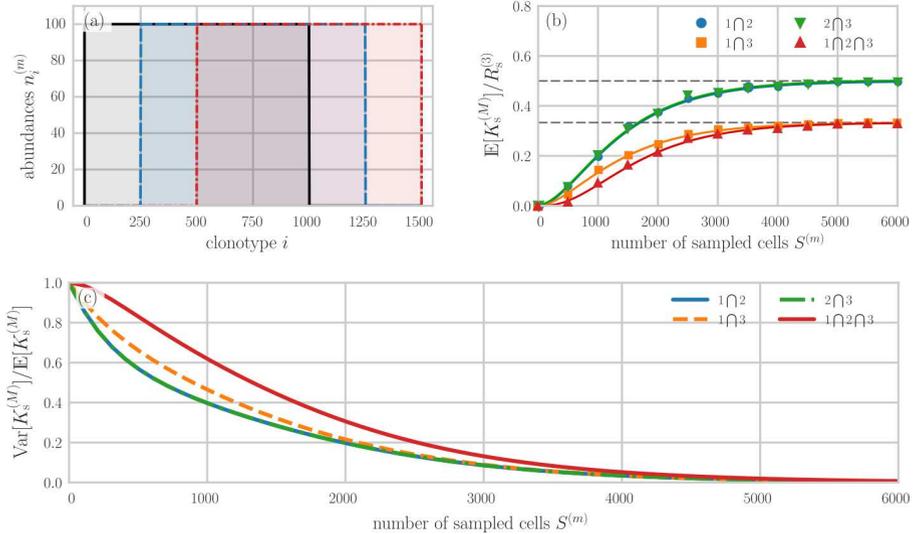}
\caption{Sampling from shifted uniform distributions. (a) Synthetic
  TCR or BCR distributions of $M=3$ individuals. The distributions in
  individuals 1, 2, and 3 are indicated by solid black, dashed blue,
  and dash-dotted red lines, respectively. Each individual has
  $10^{5}$ cells uniformly distributed across $1000$ clones ($100$
  cells per clone). The sampled group richness $R_{\rm s}^{(3)}$ is
  $1500$. (b) Samples of size $S^{(m)}$ have been generated to compute
  the relative overlaps between individuals 1 and 2 (blue disks), 2
  and 3 (green inverted triangles), 1 and 3 (orange squares), and 1--3
  (red triangles). The solid black lines show the corresponding
  analytical solutions $\mathds{E}\big[K_{\rm s}^{(M)}\big]/R_{\rm
    s}^{(3)}$ (see Eq.~\eqref{K_M_p}). Dashed grey lines show the
  maximum possible relative overlaps $500/1500\approx 0.33$ and
  $750/1500=0.5$. (c) The Fano factor ${\rm Var}\big[K_{\rm
      s}^{(M)}\big]/{\mathbb{E}}\big[K_{\rm s}^{(M)}\big]$ associated
  with relative overlaps between individuals 1 and 2 (solid blue
  line), 2 and 3 (dash-dotted green line), 1 and 3 (dashed orange
  line), and 1--3 (solid red line) as a function of the number of
  sampled cells $S^{(m)}$.}
\label{fig:overlap}
\end{figure}

\section{Simulations}
\label{sec:simus}
\begin{figure}[htb]
\centering
\includegraphics[width=4.7in]{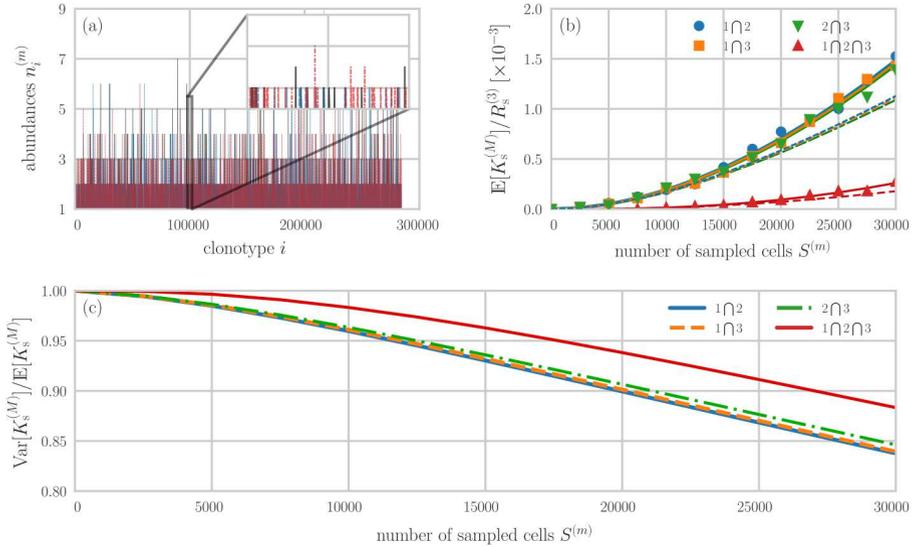}
\caption{Sampling from empirical TRB CDR3 distributions and
    overlap measures in the number representation. (a) Distributions of
    TRB CDR3 cells in $M=3$ individuals. We used the \textsc{SONIA}
    package~\cite{elhanati2014quantifying} to generate $10^{5}$ TRB
    CDR3 sequences for each individual. The sampled group richness
    $R_{\rm s}^{(3)}$ was found to be $284,598$. Equal sample sizes $S^{(m)}$
    were then drawn. (b) Relative overlaps between individuals 1 and 2
    (blue disks), 2 and 3 (green inverted triangles), 1 and 3 (orange
    squares), and 1--3 (red triangles). The solid black lines plot the
    corresponding analytical solutions $\mathds{E}\big[K_{\rm
        s}^{(M)}(\{\n^{(m)}\})\big]/R_{\rm s}^{(3)}$ found in
    Eqs.~\eqref{K_S_M_n}. The dashed curve corresponds to using using
    the estimator $\hat{p}_{i}^{(m)}= n_{i}^{(m)}/N^{(m)}$ in the
    expression $\mathds{E}\big[K_{\rm
        s}^{(M)}(\{\p^{(m)}\})\big]/R_{\rm s}^{(3)}$ (Eq.~\eqref{K_M_p}
    evaluated using $\rho_{i}^{(m)}(S)$ from Eq.~\eqref{rhoS}). (c)
    The Fano factor ${\rm Var}\big[K_{\rm
        s}^{(M)}\big]/{\mathbb{E}}\big[K_{\rm s}^{(M)}\big]$
    associated with relative overlaps between individuals 1 and 2
    (solid blue line), 2 and 3 (dash-dotted green line), 1 and 3
    (dashed orange line), and 1--3 (solid red line) as a function of
    the number of sampled cells $S^{(m)}$.}
\label{fig:overlap_empirical}
\end{figure}

The sampling theory derived in the previous sections is useful for
understanding the effect of different sampling distributions on
measurable quantities such as the proportion of shared TCRs and BCRs
among different individuals. \textbf{Figures~\ref{fig:overlap} and
  \ref{fig:overlap_empirical} show two examples of receptor
  distributions, along with the respective relative overlaps and Fano
  factors, for three individuals. To illustrate our methodology
  clearly and concisely, we utilize three shifted uniform
  distributions as models of synthetic sequence distributions in
  Fig.~\ref{fig:overlap}.} In this example, the number of TCR or BCR
sequences per individual is $N^{(m)}=10^{5}$, $(m=1,2,3)$, and the
sampled group richness $R_{\rm s}^{(3)}=1500$. \textbf{Based on the
  abundance curves shown in Fig.~\ref{fig:overlap}(a), we can readily
  obtain the overlaps between individuals 1--3 (solid black, dashed
  blue, and dash-dotted red lines), as well as between all pairs of
  individuals. The maximum possible overlap, normalized by $R_{\rm
    s}^{(3)}$, between all three individuals and between individuals 1
  and 3 is $500/1500\approx 0.33$. For the two remaining pairs, the
  corresponding maximum relative overlap, normalized by the richness
  associated with all three sampled individuals, is $750/1500=0.5$.}

Using $S^{(m)}< N^{(m)}$ sampled cells from each individual, we
observe in Fig.~\ref{fig:overlap}(b) that the increase of
$\mathds{E}[K_{\rm s}^{(M)}]/R_{\rm s}^{(3)}$ with $S^{(m)}$ is
well-described by Eq.~\eqref{K_M_p}. In Fig.~\ref{fig:overlap}(c), we
plot the Fano factor ${\rm Var}\big[K_{\rm
    s}^{(M)}\big]/{\mathbb{E}}\big[K_{\rm s}^{(M)}\big]$ as a function
of the number of sampled cells $S^{(m)}$ from each individual. For
sample sizes of about 1000 (\ie, about 1\% of the total sequence
population), the Fano factor is between 0.4 (for overlaps between
individuals 1 and 2 and between individuals 2 and 3) and 0.6 (for the
overlap between individuals 1--3). As sample sizes reach about 5\% of
the total number of sequences ($10^{5}$), the variance ${\rm
  Var}\big[K_{\rm s}^{(M)}\big]$ becomes negligible with respect to
the expected overlap $\mathbb{E}\big[K_{\rm s}^{(M)}\big]$.

As an example of an application to empirical TRB CDR3 data, we used
the \textsc{SONIA} package~\cite{elhanati2014quantifying} to generate
amino acid sequence data for three individuals, each with
$N^{(m)}=10^5$ cells. The combined richness across all individuals is
$R_{\rm s}^{(3)}=284,598$. We show the abundances of all sequences in
Fig.~\ref{fig:overlap_empirical}(a). The majority of sequences
  has an abundance of 1 while only very few sequences have abundances
  that exceed 5. Figure~\ref{fig:overlap_empirical}(b) shows the
  expected number of shared sequences as a function of the sampled
  number of cells $S^{(m)}$. To evaluate Eq.~\eqref{K_S_M_n}, we
  compute the binomial terms in $\tilde{\sigma}_{i}$ and
  $\tilde{\sigma}_{ij}$ by expanding them according to,
  \textit{e.g.},

\begin{equation}
\displaystyle  \frac{{N^{(m)}-n_{i}^{(m)}\choose S^{(m)}}}{{N^{(m)}\choose S^{(m)}}} =
  \prod_{\ell=1}^{n_{i}^{(m)}}\left(1-\frac{S^{(m)}}{N^{(m)}-n_{i}^{(m)}+\ell}\right),
\end{equation}
where $S^{(m)}/N^{(m)}$ is the sample fraction drawn from the $m^{\rm
  th}$ individual. For large $n_{i}$, other approximations, including
variants of Stirling's approximations can be employed for fast and
accurate evaluation of binomial terms.

We compare these number-representation results with the
$\p$-representation results by using the estimates $\hat{p}_{i}^{(m)}
= n^{(m)}_{i}/N^{(m)}$ in $\rho_{i}^{(m)}(S)$ and $\rho_{ij}^{(m)}(S)$
to compute the quantities in Eqs.~\eqref{K_M_p} and \eqref{K_M_p_2}.
If the number of sampled cells $S^{(m)}$ is not too large, the
analytic approximation of using $\hat{p}_{i}^{(m)}$ in
$\rho^{(m)}_{i}(S)$ to calculate $\mathds{E}[K_{\rm s}^{(M)}]/R_{\rm
  s}^{(3)}$ is fairly accurate, as shown by the dashed back curves in
Fig.~\ref{fig:overlap_empirical}(b). Since the abundances of the
majority of sequences are very small, finite-size effects lead to
deviations from the naive approximation \eqref{rhoS} as the numbers of
sampled cells $S^{(m)}$ grows large. Of course, we can also extract
generation probabilities from \textsc{SONIA} and directly use
Eq.~\eqref{K_M_p} and $\rho_{i}^{(m)}(S)$ from Eq.~\eqref{rhoS} to
find the $\p$-representation $M$-overlap $\mathds{E}\big[K_{\rm
    s}^{(M)}(\{\p^{(m)}\})\big]/R_{\rm s}^{(3)}$.
    
To examine the variance associated with a given expected number of
shared empirical TRB CDR3 sequences, we show the Fano factor ${\rm
  Var}\big[K_{\rm s}^{(M)}\big]/{\mathbb{E}}\big[K_{\rm s}^{(M)}\big]$
as a function of the sample size $S^{(m)}$ in
Fig.~\ref{fig:overlap_empirical}(c). For the shown sample sizes up to
$S^{(m)}=3\times 10^{4}$, the Fano factor is larger than about 0.85,
indicating a relatively large variance ${\rm Var}\big[K_{\rm
    s}^{(M)}\big]$. In addition to reporting mean values of measures
of sequence sharing (\ie, ``overlap'' or ``publicness'') when
analyzing empirical receptor sequence
data~\cite{elhanati2018predicting,ruiz2023modeling}, we thus recommend
to compute ${\rm Var}\big[K_{\rm s}^{(M)}\big]$ to determine
corresponding confidence intervals.

Calculations were performed on an AMD\textsuperscript{\textregistered}
Ryzen Threadripper 3970 using \texttt{Numba} to parallelize the
calculation of Eqs.~\eqref{K_S_M_n} and $\sum_{j\neq
  i}^{\Omega}\prod_{m=1}^{M}\rho_{ij}^{(m)}$ used in ${\rm
  Var}\big[K^{(M)}\big]$.

\section{Explicit forms for power-law probabilities}
\label{sec:forms}

All of our results thus far assume a model or estimate of $p_{i}$ or
$n_{i}$, as well as knowledge of at least $\Omega$. For our formulae
to be useful, the theoretical maximum richness $\Omega$ also needs to
be estimated or modeled.  Numerous parametric and nonparametric
approaches have been developed in the statistical ecology literature
\cite{CHAO1992,WANG2005,GOTELLI2011,COLWELL2012,GOTELLI2013,CHAO2014,CHAO,CHAO2020},
as well as expectation maximization methods to self-consistently
estimate richness and most likely clone population $\n$
\cite{kaplinsky2016robust}.

\begin{figure}[t]
\centering
\includegraphics[width=4.7in]{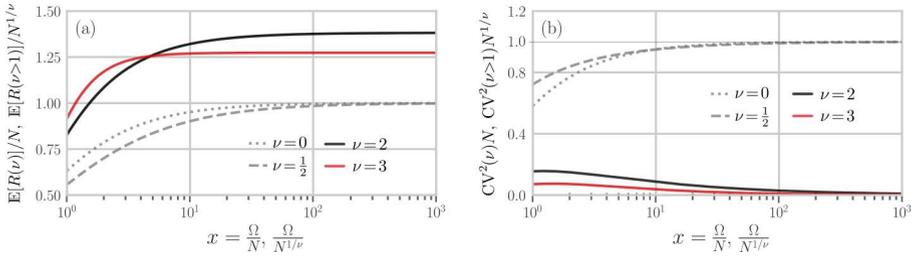}
\caption{Expected richness and uncertainty under power law-distributed
  probabilities $p_{i}$ following Eq.~\eqref{Pi}. (a) Expected
  richness for different values of $\nu$ that lead to simple scaling
  and dependence only on $x=\Omega/N, \Omega/N^{1/\nu}$.  For large
  $x$, the expected rescaled richnesses $\mathds{E}[R(\nu)]/N$ and
  $\mathds{E}[R(\nu>0)]/N^{1/\nu}$ converge. Since the normalization
  of the expected richness (by $N^{1/\nu}$) for $\nu > 1$ is different
  than for $\nu=0,\tfrac{1}{2}$, (normalized by $N$),
  $\mathds{E}[R(\nu>0)]/N^{1/\nu}$ converges to greater values, but
  $\lim_{x\to \infty}\mathds{E}[R(\nu>0)]$ remains $<1$.  (b) From the
  variances $\mathds{E}[R^{2}(\nu)]$, we construct the squared
  coefficient of variation and plot ${\rm CV}^{2}N \equiv N{\rm
    var}[R(\nu)]/(\mathds{E}[R(\nu)])^{2}$ as a function of
  $x=\Omega/N$ (or ${\rm CV}^{2}(\nu > 1)N^{1/\nu}$ for as a function
  of $x=\Omega/N^{1/\nu}$). For large $x$, ${\rm
      CV}^{2}\approx 1/N$ for $\nu = 0,1/2$ but ${\rm
      CV}^{2}N^{1/\nu}\sim 0$ for $\nu =2,3$.}
 \label{POWER_PLOT}
\end{figure}

To explore how our results depend on parameters such as $\Omega$ we
derive approximate analytic expressions when the \textbf{identical
  individual probabilities $p_{i}^{(m)}=p_{i}$} obey truncated
power-law distributions:

\begin{equation}
    p_{i} = \frac{i^{-\nu}}{H_{\nu}(\Omega)},\quad  p_{j} \leq p_{i}\,\,\,
    \mbox{if }\nu \geq 0, \,\, j\leq i, \,\,\, i,j = 1,2,\ldots, \Omega,
    \label{Pi}
    %\quad i = 1,2,\ldots,\Omega,
     \end{equation}
where $H_{\nu}(\Omega)\equiv \sum_{j=1}^{\Omega} j^{-\nu}$.  If
$\Omega$ is sufficiently large, we would like to show under what
conditions the expectations of our diversity measures converge quickly
to $\Omega$-independent values.  By approximating
$\sum_{i=1}^{\Omega}(1-p_{i})^{N} \approx \sum_{i=1}^{\Omega}
e^{-Np_{i}} \approx \int_{1}^{\Omega}
e^{-N/(H_{\nu}(\Omega)z^{\nu})}\dd z$ in Eq.~\eqref{rhoN} we find in
the large $\Omega$ limit

%For example, using $p_{i}$ in, say,
%Eqs.~\eqref{rhoN} and \eqref{R_MOMENT1}, the expected richness can be
%accurately approximated by

  \begin{equation}
    \begin{aligned}
      \frac{\mathds{E}[R(\nu=0)]}{N} \approx &\, x(1-e^{-1/x}), & x \equiv \Omega/N\qquad  & \:  \\
 \frac{\mathds{E}[R(\nu=\tfrac{1}{2})]}{N} \approx &\, 
x\left[1-2\mbox{E}\big(3,\tfrac{1}{2x}\big)\right], & x \equiv \Omega/N \qquad  & \: \\
\frac{\mathds{E}[R(\nu=1)]}{N} \approx &\, 1-\frac{\log N}{\log \Omega}+
\frac{\log(\log \Omega)}{\log \Omega}, & \:  & \: \\
 \frac{\mathds{E}[R(\nu>1)]}{N^{1/\nu}}  \approx &\, x\left[1-\tfrac{1}{\nu}\mbox{E}\big(1+\tfrac{1}{\nu}, \tfrac{x^{-\nu}}{\zeta(\nu)}\big)\right], & x\equiv \Omega/N^{1/\nu} \,\, & \:
%
%
%      \mbox{Var}[R(\p)] \approx  & \Omega e^{-N/\Omega}\big(1-e^{-N/\Omega}\big) \approx 
%      N\left[1-\tfrac{3N}{2\Omega}+\tfrac{7}{6}\left(\tfrac{N}{\Omega}\right)^{2}
%        +\ldots\right],
\label{R_NU}
 \end{aligned}
  \end{equation}
where the exponential integral is defined by $\mbox{E}(x,y)\equiv
\int_{1}^{\infty}\!t^{-x}e^{-yt}\dd t$ and $\zeta(\nu)$ is the Riemann
zeta function.  Consistent with known biology and previous
estimates~\cite{zarnitsyna2013estimating,lythe2016many}, we take the
large-$\Omega$ limit where $x > 1$.  From Eqs.~\eqref{R_NU}, we see
that the expected richnesses converge to fixed values for large enough
$\Omega$ and all values of $\nu \not\approx 1$. The rescaled expected
richnesses are plotted as functions of $x=\Omega/N$ or
$x=\Omega/N^{1/\nu}$ in Fig.~\ref{POWER_PLOT}(a).

%\begin{equation}
%  \mathds{E}[R(\p)] \approx \left({N \over Z_{\Omega}(\nu)}\right)^{1/\nu}
%  \tfrac{1}{\nu} \big\vert\Gamma\big(-\tfrac{1}{\nu}\big)\big\vert + O\big(N/\Omega^{\nu
%  - 1}\big)
%\end{equation}
%
%in the $\Omega \gg \big(N/Z_{\Omega}(\nu)\big)^{1/\nu}\gg 1$ limit,
%where $Z_{\Omega\gg 1}(\nu) \approx \zeta(\nu)$ can also be
%approximated by the $\zeta$-function. 

%If the truncation of probabilities is increased from $\Omega$ to
%$\Omega_{0} > \Omega$ are included in the summation, the expected
%richness takes on an additional term proportional to
%
%\begin{equation}
%\sum_{i=\Omega+1}^{\Omega_{0}}\big[1-(1-i^{-\nu}/Z_{\Omega_{0}(\nu)})^{N}\big]
%  \sim \left({N\over \zeta(\nu)\Omega_{0}^{\nu-1}}\right)
%  \Big[(\Omega_{0}/\Omega)^{\nu-1}-1\Big]
%\end{equation}
%
%which can also be negligible provided $\nu$ is sufficiently large.

Analogous cutoff-insensitive results can be found for the variance
${\rm Var}[R^{2}(\nu)]$ as well as other quantities. A good
approximation for the variance is

\begin{equation}
\begin{split}
\mbox{var}[R(\nu)] &  = \mathds{E}\big[R^{2}(\nu)\big] -  \big(\mathds{E}[R(\nu)]\big)^{2} \\
\: & \approx \sum_{i=1}^{\Omega} e^{-p_{i}N}\big(1-e^{-p_{i}N}\big) \\
\: &  \approx \tfrac{\Omega}{\nu}\left(\mbox{E}\big(1+\tfrac{1}{\nu}, \tfrac{N}{H_{\nu}(\Omega)\Omega^{\nu}}\big) - \mbox{E}\big(1+\tfrac{1}{\nu}, 
\tfrac{2}{H_{\nu}(\Omega)\Omega^{\nu}}\big)\right),
\end{split}
\end{equation}
where the power-law assumption in Eq.~\eqref{Pi} is used in the last
approximation. The normalized squared coefficients of variation
$\mbox{CV}^{2} \equiv \mbox{var}[R(\nu)]/(\mathds{E}[R(\nu)])^{2}$ for
representative $\nu$ are found to be

  \begin{equation}
    \begin{aligned}
    {\rm CV}^{2}_{\nu =0}N \approx & \, \frac{e^{-1/x}}{x(1-e^{-1/x})},
%\approx 1- \frac{1}{2x}+{\cal O}(x^{-3}), 
& x \equiv \Omega/N\qquad\quad  & \:  \\
{\rm CV}^{2}_{\nu =1/2}N \approx & \, \frac{2}{x}
\frac{{\rm E}(3,\tfrac{1}{2x}) -{\rm E}(3,\tfrac{1}{x})}{\left(1-2{\rm E}(3,\tfrac{1}{2x})\right)^{2}},
%\approx 1- \frac{\log x}{4x}+{\cal O}(x^{-1}), 
& x \equiv \Omega/N\qquad\quad  & \:  \\
{\rm CV}^{2}_{\nu =1}N  \approx & \, \frac{\log \Omega \big[\log(\Omega\log \Omega)-\log 4N\big]}{\big[\log(\Omega\log\Omega)-\log N\big]^{2}}, & \:  & \:  \\
{\rm CV}^{2}_{\nu > 1}N^{1/\nu}  \approx & \, \frac{1}{\nu x}\frac{{\rm E}\big(1+\tfrac{1}{\nu},
\tfrac{x^{-\nu}}{\zeta(\nu)}\big) -{\rm E}\big(1+\tfrac{1}{\nu},
\tfrac{2x^{-\nu}}{\zeta(\nu)}\big)}{\left(1-\tfrac{1}{\nu}{\rm E}\big(1+\tfrac{1}{\nu},
\tfrac{x^{-\nu}}{\zeta(\nu)}\big)\right)^{2}}, & 
x \equiv \Omega/N^{1/\nu} \qquad  & \: 
\label{R_CV}
 \end{aligned}
  \end{equation}
Plots of the CV of the richness under power-law system probabilities
are shown in Fig.~\ref{POWER_PLOT}(b). We see that the squared CVs
converge in the large $x$ limit to $N^{-1}$ for $\nu =0,1/2$ and
vanish for $\nu > 1$.

%%%%%%%%%%%%%%%%%%%%%%%%%%%%%%%%%%%%%%%%%%%%%%%%%%%%%%%%%%%
%%%%%%%%%%%%%%%%%%%%%%%%%%%%%%%%%%%%%%%%%%%%%%%%%%%%%%%%%%%

\textbf{The behavior of the sampled $M$-overlap,
  $\mathds{E}[K_{\rm s}^{(M)}(\{\p^{(m)}\})]$, can also be quantified
  under the power-law probability distribution.  By using
  Eq.~\eqref{K_M_p} and $\rho_{i}(S)$ (Eq.~\eqref{rhoS}, assuming
  equal probabilities $p_{i}^{(m)}=p_{i}$ and sample sizes $S^{(m)}=S$
  across individuals), we find}
\begin{equation}
   \mathds{E}\big[K_{\rm s}^{(M)}(\{\p^{(m)}\})\big] =  
\sum_{i=1}^{\Omega}\prod_{m=1}^{M}\rho_{i}^{(m)}(S) \approx 
\sum_{i=1}^{\Omega}\big(1-e^{-p_{i}S}\big)^{M}.
\end{equation}
\textbf{This expression can be further simplified in the large $\Omega$ limit
for specific $\nu$,}

\begin{equation}
\begin{aligned}
\mathds{E}\big[K_{\rm s}^{(M)}(\nu=0)\big] & =
\Omega\big(1-(1-1/\Omega)^{S}\big)^{M} \approx \Omega
\big(1-e^{-S/\Omega}\big)^{M}, \\
%
%\mathds{E}\big[K_{\rm s}^{(M)}(\nu \lesssim 0.5)\big] & =\tfrac{1}{\nu}
%\left(\frac{S}{H_{\nu}(\Omega)}\right)^{M}\left[\Omega^{1-M\nu}\mbox{E}\big(
%\\
%
\mathds{E}\big[K_{\rm s}^{(M)}(\nu \gtrsim 0.7)\big] & =
\sum_{i=1}^{\Omega}\big(1-(1-p_{i})^{S}\big)^{M} \approx
\sum_{i=1}^{\Omega}\exp\big[-Me^{-\frac{Si^{-\nu}}{H_{\nu}(\Omega)}}\big]
\\ \: & \sim\left[1-e^{-\frac{S}{H_{\nu}(\Omega)}}\right]^{M}
\left(\frac{S}{H_{\nu}(\Omega)\log M}\right)^{1/\nu}\!\!\!\!\!, \quad
S, \frac{S}{H_{\nu}(\Omega)}\gg 1.
\end{aligned}
\label{EK}
\end{equation}
\textbf{The last approximation is most accurate for $\nu > 1$ where
$H_{\nu}(\Omega\to \infty)$ converges and the prefactor in brackets is
$\approx 1$. For sufficiently large $S$, it still provides a rough
estimate of $M$-overlap for smaller values of $\nu$.
Asymptotic expressions for even smaller values of $\nu$ can be found
in the $S/H_{\nu}(\Omega) \ll 1$ limit, but this limit yields
very low expected $M$-overlap and is typically less informative.}
\begin{figure}[t]
\centering
\includegraphics[width=4.6in]{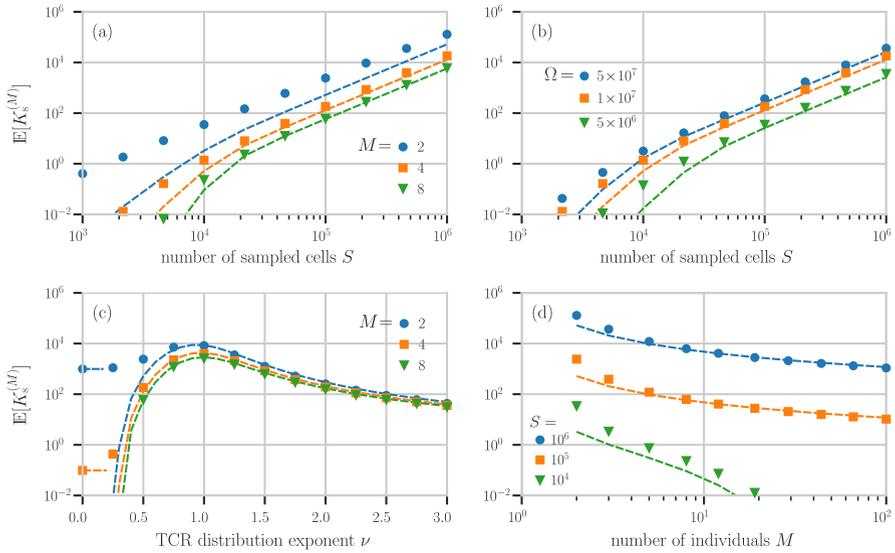}
\caption{\textbf{The expected $M$-overlap $\mathds{E}[K_{\rm
        s}^{(M)}(\{\p^{(m)}\})]$. (a) Log-log plot of $M$-overlap as a
    function of individual sample size $S$ using $p_{i} =
    i^{-1/2}/H_{1/2}(\Omega)$ ($\nu=1/2$) and $\Omega =
    10^{7}$. $M=2,4,8$ are shown, with exponentially decreasing
    $M$-overlap as $M$ is increased. (b) Fixing $M = 4$, a log-log
    plot of $\mathds{E}\big[K_{\rm s}^{(4)}(\nu=1/2)\big]$ against $S$
    for different values of $\Omega= 5\times 10^{6}, 10^{7}$ and
    $5\times 10^{7}$. (c) $\mathds{E}\big[K_{\rm s}^{(M)}(\nu)\big]$
    plotted against $\nu$ for fixed $\Omega = 10^{7}$ and different
    $M$. (d) With $\Omega=10^{7}$, a log-log plot of
    $\mathds{E}\big[K_{\rm s}^{(M)}(\nu=1/2)\big]$ as a function of
    the number $M$ of individuals sampled with $S=10^4, 10^{5},
    10^{6}$. In all panels, the dashed curves plot the analytic
    approximation for $\nu
    \gtrsim 0.7$ given in the second line of \eqref{EK}. In (c), the
    $\nu=0$ limit matches the expression given by the first line in
    \eqref{EK}. The approximations given in \eqref{EK} are especially
    accurate for large $S$ and larger $M$ and $\nu$.}}
 \label{fig:EK}
\end{figure}

\textbf{Fig.~\ref{fig:EK} plots the $M$-overlap $\mathds{E}\big[K_{\rm
      s}^{(M)}(\nu)\big]$ as a function of sample size, power-law
  $\nu$, and $M$. For comparison, the analytic approximation for $\nu \gtrsim 0.7$
  \eqref{EK} is also plotted by the dashed curves. Eq.~\eqref{EK} and
  plots such as those in Fig.~\ref{fig:EK}(a,b) could be useful for
  estimating the sample size $S$ required in order to observe a
  specific overlap between the immune repertoires of $M$ selected
  individuals. For instance, with $M=4$ individuals, a repertoire size
  of $\Omega=10^7$, and a sequence distribution exponent $\nu=0.5$, an
  expected $M$-overlap of approximately 1 can be achieved with a
  sample size of $S=10^4$.}

\textbf{Since the $\big(1/H_{\nu}(\Omega)\big)^{1/\nu}$ term in
    Eq.~\eqref{EK} increases with $\nu$, we expect that an effectively
    smaller repertoire size (recall $p_{i} \sim i^{-\nu}$ and larger
    $\nu$ leades to fewer larger-population clones), that the expected
    $M$-overlap increases with $\nu$.  However, the $\big(S/\log
    M\big)^{1/\nu}$ factor decreases with $\nu$ since larger $S$ give
    rise to a larger number of ways clones sampled from different
    individuals can ``avoid'' each other. These features give rise to
    a maximum in $\mathds{E}\big[K_{\rm s}^{(M)}(\nu)\big]$, as shown
    in Fig.~\ref{fig:EK}(c).}

\textbf{Using Eqs.~\eqref{K_S_M_n} and \eqref{rhoijS}, we can
  also straightforwardly evaluate the variance of the $M$-overlap. For
  $\nu=0$ and uniform $p_{i} = 1/\Omega$, $\mbox{var}\big[K_{\rm
      s}^{(M)}(\nu=0)\big]\approx \Omega
  (1-e^{-S/\Omega})^{M}\big(1-(1-e^{-S/\Omega})^{M}\big)$.  We find
  the variances of the $M$-overlap, as with our other metrics, are
  well approximated by that of a binomial process in the $S \to
  \infty$ limit and when values of $\nu$ are modest:
  $\mbox{var}\big[K_{\rm s}^{(M)}\big]\approx
  \Omega\tfrac{\mathds{E}[K_{\rm
        s}^{(M)}]}{\Omega}\big(1-\tfrac{\mathds{E}[K_{\rm
        s}^{(M)}]}{\Omega}\big)$.
Modest deviations from this approximation arise for finite $S$ and
large values of $\nu$.}
\section{Sampling resolution and information loss}
\label{sec:information_loss}
We end with a brief discussion of information loss upon
coarse-graining which arises when analyzing lower-dimensional
experimental/biochemical classifications of clones that are commonly
used. Such lower-dimensional representations can be obtained through
spectratyping~\cite{gorski1994circulating,fozza2017study}. For TCRs,
spectratyping groups sequences together and produces compressed
receptor representations describing CDR3 length, frequency, and
associated beta variable (TRBV) genes~\cite{gkazi2018clinical}. In
addition to coarse-grained representations of sequencing data, some
studies~\cite{elhanati2018predicting,ruiz2023modeling} use
continuous approximations to describe the distribution of receptor
sequences. Estimators of entropy and their errors have been developed
for subsampling from discrete distributions
\cite{SCHURMANN_2004,GRASSBERGER_2022}. Therefore, in this section, we
focus on quantifying differences in information content that are
associated with (i) using continuous approximations of discrete
sequencing data, and (ii) coarse-graining already-discretized (\ie,
spectratyping) distributions.

% or estimates of subsampled distributions
%  \cite{SCHURMANN_2004,GRASSBERGER_2022}.}

Given a discrete random variable $X$ describing $\Omega$ ``traits''
and taking on possible values $\{x_1,x_2,\dots,x_{\Omega}\}$, let
$p_i=\mathds{P}(X=x_i)$. The entropy of this probability distribution
is given by $H_p=-\sum_{i=1}^{\Omega} p_i \log(p_i)$.  Similarly, one
might define the differential entropy for a continuous random variable
taking on values in the interval $[a,b]$ as $S_{p} = -\int_a^b p(x)
\log p(x) \ {\rm d} x$.  It is well-known that the differential
entropy is not a suitable generalization of the entropy concept to
continuous variables~\cite{JAYNES1963} since it is not invariant under
change of variables and can be negative.  These issues can be
circumvented by introducing the limiting density of discrete
points. Here, we present a more direct approach that will be
sufficient for our application. For a probability density function
$p:\ [a,b] \to \mathbb{R}_0^+$ we introduce a discretizing morphism
$\mathcal{D}_{\Delta}$ so that
\begin{align}
	q_i = \int_{a+(i-1)\Delta}^{a+i\Delta} p(x) {\rm d} x, 
	\quad i = 1,2,\dots, B.
\end{align}
describes a random variable taking on values in each of the
$(b-a)/\Delta = B$ bins.

To quantify the amount of information lost in this discretization
step, consider the entropy $H_q = -\sum_{i=1}^{B}q_{i}\log q_{i} \sim
\log \Delta$ in the $\Delta\to 0$ limit.
\begin{figure}[t]
\centering
\includegraphics{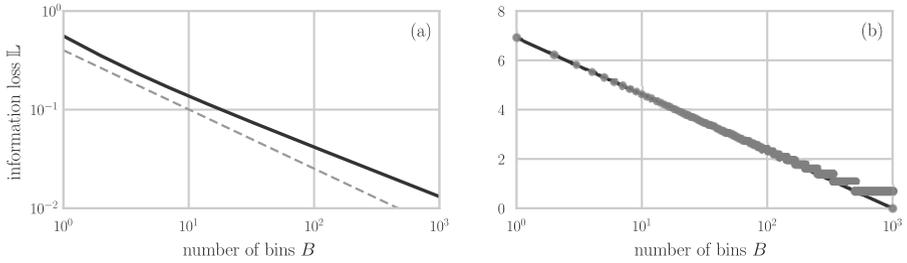}
\caption{The information loss $\mathbb{L}$ as a function of the number
  of discretization bins $B$. The loss is least as the number of
  integration bins $B\to \infty$. (a) The solid black line shows the
  information loss associated with discretizing a truncated power law
  [see Eq.~(\ref{eq:p})], and the dashed grey line is a
  guide-to-the-eye (power law) with slope $-0.6$. (b) Grey dots show
  the information loss associated with coarse graining a discrete and
  uniform random variable with initially $\Omega=1000$ traits. The
  solid curve shows the corresponding analytical result for the
  difference in information $\mathbb{L} = -\log (B/\Omega)$ between
  discretizing a continuous uniform distribution of
    $\Omega=1000$ traits using $B$ bins.}
 \label{fig:loss}
\end{figure}
If we want to evaluate any information loss as a difference between
the (finite) differential entropy $S_{p}$ and the (diverging) entropy
$H_{q}$ we need to account for this logarithmic contribution
by defining the corresponding information loss as
\begin{equation}
	\mathbb{L}(\Delta) = \vert (S_{p}- \log \Delta)  - H_q\vert.
\label{LDELTA}
\end{equation}
By absorbing the logarithmic contribution into the differential
entropy, we find the correct continuous entropy according to
Jaynes~\cite{JAYNES1963} using the limiting density of discrete
points. 

As an example, we compute the information loss associated
with discretizing the truncated power law
\begin{equation}
\label{eq:p}
p(x) = \left\{\begin{array}{ll}
	       	\frac{1}{\gamma(\tfrac{1}{2},1)}
\frac{e^{-x}}{\sqrt{x}}, & \text{if}\quad 0\leq x \leq 1
	       	\\[14pt]
	       	0, & \text{else}
	       \end{array}\right.,
\end{equation}
where $\gamma(s,x)=\int_{0}^{x} t^{s-1} e^{-t}\,\mathrm{d}t$ is the
lower incomplete gamma function. The distribution Eq.~\eqref{eq:p}
gives rise to few high-abundance clones and many low-abundance clones,
as typical for TCR receptor sequences~\cite{xu2020diversity}.
Analytic expressions for the discretized probabilities $q_i$ are
lengthy, so we numerically compute $q_i$ to evaluate the information
loss $\mathbb{L}(\Delta)$. Eq.~\eqref{LDELTA} is plotted as a function
of the number of bins $B=1/\Delta$ in Fig.~\ref{fig:loss}(a).  The
information loss decreases with the number of bins, as this results in
the discrete distribution gathering more information about its
continuous counterpart.

While the connection between continuous probability distributions and
their discretized counterparts has important consequences for
sampling, information loss also occurs in spectratyping when an
already discrete random variable is coarse grained. In this scenario,
the information loss can be quantified uniquely (up to a global
multiplicative constant) by the entropy difference of the
distributions~\cite{Baez11}. The difference between the full
$H_p=-\sum_{i=1}^{\Omega} p_i \log(p_i)$ and the coarse-grained $H_q =
-\sum_{i=1}^{B}q_{i}\log q_{i}$ can be explicitly 
evaluated for uniformly distributed probabilities.

For any number $B<\Omega$ we can define a coarse graining procedure
that yields only $B$ traits by defining the bin size
$\Delta=\operatorname{ceil}(\Omega/B)$ and grouping together $\Delta$
traits into each bin.  The last bin might be smaller than the other
bins or even empty. The information loss of this procedure is shown in
Fig.~\ref{fig:loss}(b) for an initially uniform distribution of
$\Omega = 1000$ traits. Across certain ranges of $B$, plateaus can
build since our coarse graining might add zero probabilities.
However, we can instead start from a continuous distribution and
compare the discretization with $\Omega =1000$ bins to any other
binning with $B\leq \Omega$.

Comparing a coarse-grained uniform distribution with $B$ bins to the
discretized distribution with $\Omega$ bins yields the information
loss with respect to the initial discrete distribution $\mathbb{L} =
-\log (B/\Omega) \geq 0$.  We plot this analytical prediction against
the information loss $\mathbb{L}$ associated with coarse graining an
already discrete distribution in Fig.~\ref{fig:loss}(b), showing them
to be well-aligned.

\section{Discussion and conclusions}
\label{sec:conclusion}
Quantifying properties of cell-type or sequence distributions is an
important aspect of analyzing the immune repertoire in humans and
animals. Different methods have been developed to estimate TCR and BCR
diversity indices such as the total number of distinct sequences in an
organism (\ie, species richness)
\cite{rempala2013methods,kaplinsky2016robust,xu2020diversity}. Another quantity of
interest is the number of clones that are considered ``public'' or
``private,'' indicating how often certain TCR or BCR sequences occur
across different individuals.

Public TCR$\beta$ and BCR sequences have been reported in a number of
clinical
studies~\cite{putintseva2013mother,robins2010overlap,shugay2013huge,
  soto2020high,briney2019commonality,soto2019high}. However, the terms
``public'' and ``private'' clonotypes are often based on different and
ambiguous definitions. According to \cite{shugay2013huge}, a ``public
sequence'' is a sequence that is ``\emph{often} shared between
individuals''~\cite{shugay2013huge}, while \cite{greiff2017learning}
refers to a sequence as public if it is ``shared across
individuals''. In addition to ambiguities in the definition of what
constitutes a private/public sequence, overlaps between the immune
repertoires of different individuals are often reported without
specifying confidence intervals, even though variations may be large
given small sample sizes and heavy tailed sequence distributions.

\textbf{In this work, we provided mathematical definitions for
  ``public'' and ``private'' clones in terms of the
  probabilities of observing a number of clones across $M$ selected individuals, complementing related work that introduced the
  notion of ``sharing number $M$'' (\ie, the expected number of
  sequences which will be found in \emph{exactly} $M$ individuals) to
  quantify the expected overlap between cell-sequence
  samples~\cite{elhanati2018predicting,ruiz2023modeling}.}
\textbf{Besides defining individual repertoire probability
  distributions, our results include analytic expressions for
  individual and multi-individual expected richness and expected
  overlap as given by Eqs.~\eqref{eq:richness_one}, \eqref{R_MOMENT1},
  \eqref{RM_AVE}, \eqref{KM}, and \eqref{K_M_p}. Additionally, using
  Eqs.~\cref{RICHNESS_S,RICHNESS_S_M,K_S_M,R_n,RICHNESS_S_M_n,K_S_M_n}, we derived
  expressions for the expected richness and expected overlap in
  subsamples. The variability of quantities (second moments) such as
  the $M$-overlap and subsampled overlap were also derived. Studies
  analyzing the similarities and differences associated with immune
  repertoires of different individuals (see, \eg,
  \cite{elhanati2018predicting,soto2020high,ruiz2023modeling}) may
  utilize our results on second moments of overlap measures to
  quantify the statistical significance of their findings.} Our
results are summarized in Table \ref{tab:results} where we provide
expectations and second moments of all quantities as a function the
cell population configurations $\n^{(m)}$ or as a function of the
underlying clone generation probabilities $\p^{(m)}$, as is generated
by models such as \textsc{SONIA}~\cite{elhanati2014quantifying}.

%These quantities can also be straightforwardly computed in the
%``clone count'' $\c$-representation, which we leave as an exercise
%for the reader.

Further inference of richness and overlap given sample configurations
can be developed using our results. For example, the parametric
inference of expected richness in an individual given a sampled
configuration $\s$ can be found using the multinomial model and Bayes'
rule, as presented in Eq.~\eqref{INFER_R}. 

%is difficult to estimate, our results can be fairly insensitive to
%$\Omega$. This can be explicitly shown under certain truncated
%power-laws for $p_{i}$:

%{\color{red} This should be OK. Just wanted to add a note for you: In some
%simulations, I have observed that the majority of sequences only occur
%once. So, I think that in some cases, the decay of the sequence
%distribution function is very slow. In such cases, it will still be
%difficult to estimate $\Omega$. However, as shown in section 5, we can
%establish that $\Omega$ doesn't matter so much for at least some decay
%exponents.}

While our results depend on knowledge of $\Omega$ and $N$, we show
using power-law probability distributions and explicit expressions in
Eqs.~\eqref{R_NU} and \eqref{R_CV} that the richness is insensitive to
$\Omega$ in the large $N$ and $\Omega/N$ limits.  Therefore, even
though $\Omega$ may be impossible to accurately estimate, power-law
probability distributions generally render our results robust to
uncertainty in $\Omega$. Analytic or semi-analytic expressions for the
overlap quantities can also be derived. We leave this exercise to the
reader.

Finally, in the context of coarse-graining, or spectratyping
\cite{CIUPE2013}, we have discussed methods that are useful to
quantify the information loss associated with different levels of
coarse graining TCR and BCR sequences. The results presented here are
based on an assumption of simple multinomial distributions as the
underlying population model. As we mentioned, a number of
mechanistically more realistic probability distributions have been
derived for neutral, noninteracting clone populations in steady state
\cite{BDI}. These include log series and negative binomial
distributions each requiring tailored calculations for the
corresponding richness and overlap.

%\hl{[Is there something specific that we can say about future work? I
%    don't think that clinicians need to calculate information losses?
%    Our methods are probably useful to evaluate different sequencing
%    techniques in more detail?]}

\begin{table*}[h!]
      \caption{\textbf{Table of mathematical results.} We list our
        main mathematical derivations and expressions for richness and
        overlap, unsampled and sampled, in both the
        $\n$-representation and the $\p$-representation. The component
        probabilities $\rho_{i}^{(m)}$, $\rho_{ij}^{(m)}$,
        $\tilde{\rho}_{i}$, $\tilde{\rho}_{ij}$, $\sigma_{i}^{(m)}$,
        $\sigma_{ij}^{(m)}$, $\tilde{\sigma}_{i}$, and
        $\tilde{\sigma}_{ij}$ are given in Eqs.~\eqref{rhoN},
        \eqref{rhoijN}, \eqref{rhoiM}, \eqref{rhoijM}, \eqref{sigma},
        \eqref{sigmaij}, \eqref{sigmaiM}, and \eqref{sigmaijM},
        respectively. The component probabilities $\rho_{i}^{(m)}(S)$,
        $\rho_{ij}^{(m)}(S)$, $\tilde{\rho}_{i}(S)$,
        $\tilde{\rho}_{ij}(S)$ associated with sampled quantities in
        the $\p$-representation are given by Eqs.~\eqref{rhoS},
        \eqref{rhoijS}, \eqref{rhoiMS}, and \eqref{rhoijMS},
        respectively.}
      \label{tab:results}
\begin{adjustbox}{angle=90}
\renewcommand{\arraystretch}{1.9}
\begin{tabular}{|l|l|l|l|} \hline
Measure & Description & $\n$-representation & $\p$-representation \\ \hline\hline
%\toprule
$\mathbb{E}[R]$ 
& individual richness
& \! $\sum_{k=1}^{N}\sum_{i=1}^{\Omega}\mathds{1}(n_{i},k)$
& \! $\sum_{i=1}^{\Omega}\rho_{i}$ \\[1pt]  \hline
$\mathbb{E}[R^{2}]$ 
& $2^{\rm nd}$ moment of richness 
&  \! $\left[\sum_{k=1}^{N}\sum_{i=1}^{\Omega}\mathds{1}(n_{i},k)\right]^{2}$ 
&  \! $\sum_{i=1}^{\Omega}\rho_{i}+\sum_{j\neq i}^{\Omega}\rho_{ij}$ \\[1pt] \hline
$\mathbb{E}[R_{\rm s}]$ 
& mean sampled richness 
& \! $\sum_{i=1}^{\Omega}\sigma_{i} = \Omega
- {1\over {N \choose S}}\sum_{i=1}^{\Omega} {N-n_{i}\choose S}$
& \! $\sum_{i=1}^{\Omega} \rho_{i}(S)$ \\[1pt] \hline
$\mathbb{E}[R_{\rm s}^{2}]$ 
&  $2^{\rm nd}$ moment, sampled richness
&  \! $\sum_{i=1}^{\Omega}\sigma_{i}+\sum_{j\neq i}^{\Omega}\sigma_{ij}$
&  \! $\sum_{i=1}^{\Omega}\rho_{i}(S)+\sum_{j\neq i}^{\Omega}\rho_{ij}(S)$ \\[1pt]  \hline
$\mathbb{E}[R^{(M)}]$ 
& mean group richness  
& \! $\sum_{k\geq 1}\sum_{i=1}^{\Omega}\mathds{1}\big(\textstyle{\sum_{m=1}^{M}}n_{i}^{(m)}, k\big)$  
& \! $\sum_{i=1}^{\Omega}\tilde{\rho}_{i}$ \\[1pt]  \hline
%
%$\Omega - \sum_{i=1}^{\Omega}\prod_{m=1}^{M}(1-p_{i}^{(m)})^{N^{(m)}}$ \\[1pt]  \hline
%& \approx \Omega - \sum_{i=1}^{\Omega} e^{-\sum_{m=1}^{M}p_{i}^{(m)}N^{(m)}}, 
%
%
$\mathbb{E}\big[(R^{(M)})^{2}\big]$ 
& $2^{\rm nd}$ mom., grp richness 
& \! $\Big[\sum_{k\geq 1}\sum_{i=1}^{\Omega}
\mathds{1}\big(\textstyle{\sum_{m=1}^{M}}n_{i}^{(m)}, k\big)\Big]^{2}$
& \! $\sum_{i}^{\Omega}\tilde{\rho}_{i} + \sum_{i\neq j}^{\Omega}\tilde{\rho}_{ij}$ \\[1pt] \hline
%
%\sum_{i=1}^{\Omega}\Big[1-\prod_{m=1}^{M}(1-p_{i}^{(m)})\Big] \\ \: &
%+ \Omega^{2}
%-(2\Omega-1)\sum_{i=1}^{\Omega}\prod_{m=1}^{M}\big(1-p_{i}^{(m)}\big)^{N^{(m)}}
%+\sum_{i\neq
%  j}^{\Omega}\prod_{m=1}^{M}\big(1-p_{i}^{(m)}-p_{j}^{(m)}\big)^{N^{(m)}}.
%
${\rm Var}[R^{(M)}]$ 
& variance of grp richness 
& \! $0$
& \!  $\sum\limits_{i\neq j}^{\Omega}\tilde{\rho}_{ij} + \left(1-\sum\limits_{i=1}^{\Omega}
\tilde{\rho}_{i}\right)\sum\limits_{i=1}^{\Omega}\tilde{\rho}_{i}$ \\[1pt] \hline
$\mathbb{E}[R_{\rm s}^{(M)}]$ 
& mean sampled grp richness & \!
$ \! \sum_{i=1}^{\Omega}\tilde{\sigma}_{i}$
%
%$\sum_{k\geq 1}\sum_{i=1}^{\Omega}\mathds{1}\big(\textstyle{\sum_{m=1}^{M}}s_{i}^{(m)}, k\big)$  
& \! $\sum_{i=1}^{\Omega}\tilde{\rho}_{i}(S)$ \\[1pt]  \hline
%& \approx \Omega - \sum_{i=1}^{\Omega} e^{-\sum_{m=1}^{M}p_{i}^{(m)}N^{(m)}}, 
%
$\mathbb{E}\big[(R_{\rm s}^{(M)})^{2}\big]$ 
& $2^{\rm nd}$ mom., sampled grp richness 
& \! $\sum_{i=1}^{\Omega}\tilde{\sigma}_{i} + \sum_{i\neq j}^{\Omega}\tilde{\sigma}_{ij}$
%
%$\Big[\sum_{k\geq 1}\sum_{i=1}^{\Omega}
%  \mathds{1}\big(\sum_{m=1}^{M}s_{i}^{(m)}, k\big)\Big]^{2}$
%
& \! $\sum_{i=1}^{\Omega}\tilde{\rho}_{i}(S)+ \sum_{i\neq j}^{\Omega}\tilde{\rho}_{ij}(S)$  \\[1pt] \hline
$\mathbb{E}[K^{(M)}]$ 
& expected $M$-overlap
& \! $\sum_{i=1}^{\Omega}\prod_{m=1}^{M}\sum_{k^{(m)}\geq 1}\!\mathds{1}(n_{i}^{(m)}\!,k^{(m)})$ 
& \! $\sum_{i=1}^{\Omega}\prod_{m=1}^{M}\rho_{i}^{(m)}$\\[1pt]\hline
$\mathbb{E}\big[(K^{(M)})^{2}\big]$  
& $2^{\rm nd}$ moment, $M$-overlap
& \! $\left[\sum\limits_{i=1}^{\Omega}\prod\limits_{m=1}^{M}\sum\limits_{k^{(m)}\geq 1}\!
  \mathds{1}(n_{i}^{(m)}\!,k^{(m)})\right]^{2}$ 
& \! $\sum\limits_{i=1}^{\Omega}\prod\limits_{m=1}^{M}\rho_{i}^{(m)}
+\sum\limits_{i\neq j}^{\Omega}\prod\limits_{m=1}^{M}\rho_{ij}^{(m)}$
\\[1pt]  \hline  
$\mathbb{E}[K_{\rm s}^{(M)}]$ 
& sampled $M$-overlap
& \!$\sum_{i=1}^{\Omega}\prod_{m=1}^{M}\sigma_{i}^{(m)}$ 
& \!$\sum_{i=1}^{\Omega}\prod_{m=1}^{M}\rho_{i}^{(m)}(S)$ \\[1pt]  \hline  
$\mathbb{E}[(K_{\rm s}^{(M)})^{2}]$ 
& $2^{\rm nd}$ mom., sampled $M$-overlap 
& \!$\sum\limits_{i=1}^{\Omega}\prod\limits_{m=1}^{M}\sigma_{i}^{(m)}(S)
+\sum\limits_{i\neq
  j}^{\Omega}\prod\limits_{m=1}^{M}\sigma_{ij}^{(m)}(S)$
& \!$\sum\limits_{i=1}^{\Omega}\prod\limits_{m=1}^{M}\rho_{i}^{(m)}(S)
+\!\sum\limits_{i\neq
  j}^{\Omega}\prod\limits_{m=1}^{M}\!\rho_{ij}^{(m)}(S)$
\\[1pt]  \hline
%\bottomrule
  \end{tabular}
\end{adjustbox}
  \end{table*}

\vspace{1em}
\noindent
\textbf{Data availability statement.} All source codes are publicly available at \url{https://gitlab.com/ComputationalScience/immune_repertoires}.

\bmhead{Acknowledgments} LB received funding from the Swiss National
Fund (P2EZP2\_191888) and the Army Research Office (W911NF-23-1-0129). TC acknowledges funding from the NIH through grant R01HL146552 and the NSF through grant DMS-1814364.
\bibliography{refs2}
%
%
%
%%%%%%%%%%%%%%%%%%%%%%%%%%%%%%%%%%%%%%%%%%%%%%%%%%%%%%%%
%%%%%%%%%%%%%%%%%%%%%%%%%%%%%%%%%%%%%%%%%%%%%%%%%%%%%%%%
%%%%%%%%%%%%%%%%%%%%%%%%%%%%%%%%%%%%%%%%%%%%%%%%%%%%%%%%
\end{document}